\documentclass[11pt]{article}

\usepackage[left=1in,right=1in,top=1in,bottom=1in]{geometry}
\usepackage[pdftex]{graphicx}
\usepackage{epstopdf}
\usepackage{pdflscape, longtable, sectsty, url, booktabs, rotating}
\def\sym#1{\ifmmode^{#1}\else\(^{#1}\)\fi}
\usepackage{setspace}
\usepackage{amsmath}
\usepackage{lscape}
\usepackage[final]{pdfpages}
\usepackage{verbatim}
\usepackage[bottom]{footmisc}

\usepackage{color, colortbl}
\usepackage{xcolor}
\definecolor{Gray}{gray}{0.9}
\definecolor{new}{rgb}{0.0, 0.87, 0.87}
\definecolor{new}{rgb}{0.96, 0.73, 1.0}
\definecolor{new}{rgb}{0.75, 0.58, 0.89}
\usepackage[sort,round]{natbib}
\usepackage{soul}
\usepackage{comment}
\usepackage[textsize=tiny]{todonotes}

\usepackage{titlesec}
\titleformat*{\section}{\large\bfseries}
\titleformat*{\subsection}{\normalsize\bfseries}
\titleformat*{\subsubsection}{\normalsize\bfseries}

\usepackage[labelfont=bf, bf]{caption}
\usepackage{subfigure}
\subfiguretopcaptrue
\usepackage{multirow}
\usepackage{tabularx}
\usepackage{array}
\usepackage[para, flushleft]{threeparttable}
\usepackage{float}
\usepackage{adjustbox}

\newcolumntype{L}[1]{>{\raggedright\let\newline\\\arraybackslash\hspace{0pt}}m{#1}}
\newcolumntype{C}[1]{>{\centering\let\newline\\\arraybackslash\hspace{0pt}}m{#1}}
\newcolumntype{R}[1]{>{\raggedleft\let\newline\\\arraybackslash\hspace{0pt}}m{#1}}

\usepackage[bookmarksnumbered=true, bookmarksdepth=1,citecolor=blue, colorlinks=true, filecolor=black,hyperfootnotes=true,linkcolor=red, pdffitwindow=true,urlcolor=blue]{hyperref}
\usepackage{url}
\usepackage{xr}

\usepackage{cleveref}

    \makeatletter
        \newcommand{\moverlay}{\mathpalette\mov@rlay}
         \newcommand{\mov@rlay}[2]{\leavevmode\vtop{\baselineskip\z@skip{}
             \lineskiplimit-\maxdimen\ialign{\hfil\ensuremath{#1##}\hfil\cr#2\cr\cr}}}
    \makeatother

\begin{document}

\begin{titlepage}
\thispagestyle{empty}
\title{\textbf{Educational Mobility Across \\ Multiple Generations in Indonesia}\thanks{We thank seminar participants at the Universidad de Navarra, the EEA 2021 conference, the 2023 AASLE conference in Taiwan, the 2024 EALE conference in Bergen and the RNIM seminar series. Financial support from MICIU/AEI (RYC2019-027614-I, CEX2021-001181-M) and from the ESRC-funded Centre for the
Microeconomic Analysis of Public Policy (ES/T014334/1) is gratefully acknowledged. The views expressed herein are those of the authors and should not be attributed to the Bank of Italy.}}

\author{{\textbf{Sarah Cattan}}\\[0.01in]\small{Institute for Fiscal Studies, HCEO and CESifo}\\[0.08in]
\large{\textbf{Antonio Dalla-Zuanna}}\\[0.01in]\small{Bank of Italy, IFS and IZA}\\[0.08in]
\large{\textbf{Jan Stuhler}}\\[0.01in]\small{Universidad Carlos III de Madrid}\\[0.08in]
\large{\textbf{Po Yin Wong}}\\[0.01in]\small{Queen Mary University of London}\\[0.08in]
}

\date{\today}
\maketitle
\begin{abstract}
\noindent Standard \textit{inter}generational measures have been shown to understate the long-run persistence of socioeconomic advantages in developed countries. We study theoretically and empirically whether this pattern extends to less developed settings, using Indonesia as a case study. Using the Indonesian Family Life Survey (IFLS) and Census data, we study multigenerational correlations in education across three generations. Contrary to previous findings, we observe greater multigenerational mobility than parent-child correlations alone would suggest. We develop a theoretical framework to highlight two key factors influencing multigenerational dynamics in developing countries: (1) financial and credit constraints, and (2) cultural norms related to marital sorting. To confirm  their relevance, we exploit regional variations in exposure to the 1997-98 Asian financial crisis and in marital customs.\
\end{abstract}

\small{
\vspace{30pt} 

}  

\end{titlepage}

\newpage
\doublespacing

\section{Introduction} \label{sec:intro}

Recent \textit{multi}generational studies find that traditional measures of social mobility, such as parent-child correlations, understate the persistence of socioeconomic advantages across generations. For example, when regressing a child's status on both parent and grandparent status, the coefficient for grandparents is typically positive (see e.g. \citealt{Lindahletal2013_0002}, \citealt{NeidhoeferStockhausen}, \citealt{colagrossi2020like}). This finding suggests that available estimates of social mobility may not tell us all that much about long-run persistence across multiple generations,  challenging common interpretations of intergenerational correlations and the mechanisms driving them. 

However, little is known about multigenerational processes for developing countries. Existing estimates, as reviewed by \cite{AndersonSheppardMonden2017} or \cite{StuhlerMultigenerationalHB}, are drawn almost exclusively from high-income countries, where data spanning multiple generations is more readily available. Important exceptions are \cite{KunduSen2023}, who examine mobility across three generations in India, and \cite{GallegosCelhay2016}, who compare multigenerational mobility in six Latin American countries. While these studies provide first empirical evidence for lower-income countries, as yet we know little about the mechanisms that matter in these contexts.

In this study, we explore \textit{why} multigenerational processes might differ in developing countries, illustrating our arguments using rich survey data spanning three generations from Indonesia. Our central argument is that the dynamics of multigenerational transmission -- particularly the relative strengths of multi- versus intergenerational correlations -- are bound to differ between the developing and developed world, as well as across different developing countries. The rationale is that structural factors, such as financial constraints or marital customs, operate differently in low-income countries than in high-income countries, but also have distinct dynamic implications. To support these arguments empirically, we exploit regional variations in sudden shifts in economic conditions and in marital customs. 

To begin, we first provide a brief overview of standard measures of multigenerational persistence. Most studies compare pairwise regression or correlation coefficients across two or more generations. A key question here is whether those correlations ``iterate'', i.e. whether the grandparent-child correlation is the product of the corresponding parent-child correlations. Alternatively, many studies report the slope coefficients from a multivariate regression that conditions on both parent's and grandparent's status. We note that there exists a direct mapping between the ``\textit{grandparent coefficient}'' from such multivariate regressions and the relative size of the bivariate coefficients, which simplifies comparisons across studies. In particular, this coefficient is positive if and only if multigenerational correlations are greater than the product of the underlying parent-child correlations. 

Building on standard models of intergenerational mobility, we next show that the sign of this grandparent coefficient depends on the relative strength of different transmission mechanisms. This matters as certain factors, such as parental resources and financial constraints, are more salient for child outcomes in developing countries (\citealt{Solis2017}, \citealt{Piraino:2021}, \citealt{mogstad2021family}). Another important factor is assortative mating. For a given level of parent-child correlations, multigenerational persistence may be higher when parents directly choose their children's spouses -- as is customary in some developing countries, while arranged marriages have become rare in high-income countries (\citealt{averett2018oxford}). 
Assortative practices also differ greatly across regions and ethnic groups within the developing world. %

One implication of these arguments is that, although parent-child mobility is comparatively low in developing countries (\citealt{Hertzetal2008}, \citealt{VanderWeideJDE24}), long-run mobility is not necessarily so. In line with this conjecture, we find that Indonesia has low parent-child mobility in education (consistent with estimates by \citealt{Hertzetal2008}, \citealt{Ahsanetal2024JEBO} and \citealt{VanderWeideJDE24}), yet multigenerational mobility appears not particularly low. In Indonesia, the grandparent coefficient becomes \textit{negative} when conditioning on both the father's and mother's education, while in high-income countries this coefficient is generally positive (\citealt{AndersonSheppardMonden2017}). Multigenerational mobility is therefore higher in Indonesia than an extrapolation from parent-child correlations would suggest, in contrast to the pattern in Europe (\citealt{colagrossi2020like}) or Latin America (\citealt{GallegosCelhay2016}). 

In our theoretical analysis, we compare the multigenerational patterns implied by different models of intergenerational transmission. We show that the classic Becker-Tomes framework (\citeyear{becker1979equilibrium,becker1986human}) can generate either a negative or positive grandparent coefficient, depending on parameterization. This stands in contrast to a commonly cited variant, described by \cite{Solon201413} and considered in \cite{Piraino:2021} and many other studies, in which the grandparent coefficient is necessarily negative -- implying high multigenerational mobility. We trace this contrast to a simplification of the model's earnings equation that replaces the original stochastic with a deterministic version. Though otherwise minor, this change greatly affects the model's multigenerational implications. The original Becker-Tomes model can rationalize different multigenerational patterns, and nests a simple latent factor model often used to explain positive grandparent coefficients. More generally, the sign of the grandparent coefficient depends on the relative strength of “direct” income effects and -- when extending the Becker-Tomes model to two parents -- the structure of assortative mating.

We provide evidence on these hypotheses in the empirical part of the paper, using Indonesia as a case study. To provide evidence on the role of financial constraints, we consider the 1997-98 Asian Financial Crisis. This crisis had a devastating effect on educational attainment in Indonesia: within a year, secondary school enrollment decreased from 59.4\% to 55.7\% for girls, with an even sharper decline for boys (\citealt{Poppele}). Educational spending fell dramatically, in particular among poorer households with young children (\citealt{Thomas2004}). While the crisis thus had an effect on intergenerational mobility, we are instead interested in its effects on multigenerational associations. In line with our theoretical prediction, we find a more negative coefficient on grandparents for those cohorts whose education was most directly impacted by the crisis. Thus, while financial constraints decrease intergenerational mobility in education -- consistent with previous research -- multigenerational correlations appear to be less affected.

To examine the role of assortative matching, we exploit the variation in marital customs across families, provinces and ethnic groups in Indonesia. Our data contain direct information on \textit{who} selected a person's spouse, allowing us to analyze individual-level variation in marital practices. The strength of multigenerational associations varies systematically with marital customs:  when a woman's family selects her spouse, the spousal correlation in education is weaker (i.e., spouses are less similar), but the correlation between a spouse and his or her parents-in-law is stronger (i.e., the spouses' parents are more similar to each other). Moreover, under such ``family-based'' assortative matching, the grandparent coefficient is more positive. These effects align with our theoretical hypothesis, but would \textit{amplify} multigenerational transmission, and therefore cannot explain why in the overall sample the grandparent coefficient is negative.

Our results contribute to three strands of the literature. First, they contribute to our understanding of social mobility in developing countries. Intergenerational (parent-child) mobility tends to be lower in the developing world (\citealt{MaozMoav1999}). Comparing 42 countries, \cite{Hertzetal2008} demonstrate that parent-child correlations in education are higher, and often much higher, in low- compared to high-income countries; \cite{VanderWeideJDE24} expand on this significantly by building a global database of educational mobility covering 153 countries. While there exists therefore ample evidence on intergenerational mobility, our results suggest that the relative strength of inter- versus multigenerational correlations differs in developing countries, such that long-run mobility may not be particularly low. 

Second, we contribute to the recent literature on multigenerational processes (e.g., \citealt{LindahlPalme2014_IGE4Generations}). This rapidly growing literature has established that multigenerational correlations tend to be higher than one would expect from an extrapolation of the available parent-child evidence (\citealt{Clark2014book}, \citealt{AndersonSheppardMonden2017}, \citealt{BaroneMocetti2016}). However, it remains unclear how those patterns should be interpreted (\citealt{StuhlerMultigenerationalHB}), and whether they vary across countries. We argue that multigenerational processes are bound to differ between the developed and developing world, as certain transmission mechanisms with distinct dynamic implications are more salient in the latter. 

Third, we contribute to the literature on social mobility in Indonesia, an especially interesting setting given its ethnic diversity and substantial public investments in education. A major school construction program in the 1970s significantly improved educational attainment and labor market outcomes, particularly for boys (\citealt{Duflo2001, AshrafBauNunnVoena2020}), with positive effects for girls in certain ethnic groups (\citealt{AshrafBauNunnVoena2020}, \citealt{Akreshetal2022}). The program particularly benefited children of less-educated parents, raising social mobility (\citealt{hertz2007school}) and improving outcomes in the next generation (\citealt{mazumder2019intergenerational}, \citealt{Akreshetal2022}).  Still, Indonesia exhibits relatively low parent-child mobility in both education (\citealt{Ahsanetal2024JEBO}; \citealt{RazaAytun2021}; \citealt{Hertzetal2008}; \citealt{VanderWeideJDE24}) and income (\citealt{Zafar2022, Sakrietal2022}).\footnote{A key bottleneck is secondary education: the increased number of boys graduating from primary school led to overcrowding that displaced girls from secondary education
(\citealt{ahsan2023public}), and the 1997 Asian financial crisis led to a large reduction in secondary school enrollment for both boys and girls (\citealt{Poppele}), especially among poorer families (\citealt{Thomas2004}).} Our study adds to this literature by highlighting the effects of marital customs and financial constraints on multigenerational, rather than just intergenerational, correlations in education.

The paper proceeds as follows: Section \ref{sec:measurement} discusses the conventional measurements of inter- and multigenerational transmission; Section \ref{sec:model} presents a theoretical framework to understand patterns in the transmission of education across two and three generations; Section \ref{sec:data} describes the Indonesian data and reports our baseline results on the estimated two- and three-generation correlation coefficients; Section \ref{sec:credit} provides empirical evidence on the credit constraints and direct income effects on persistence in Indonesia; Section \ref{sec:mating} examines the role of assortative mating practices on long-run persistence; and Section \ref{sec:conclusion} concludes.

\section{Measuring inter- and multigenerational transmission} \label{sec:measurement}

The empirical literature has adopted two types of measures to characterize
multigenerational transmission. First, one may consider pairwise regression
(or corresponding correlation) coefficients based on linear regressions such as
\begin{equation}
y_{it}=\alpha+\beta_{-k}y_{it-k}+\varepsilon_{it},\label{eq:pairwise_regression}
\end{equation}
where $y_{it}$ refers to the socio-economic status of an individual
in generation $t$ of family $i$ and $y_{it-k}$ is the status of an ancestor $k$ generations back. In our data, we can directly estimate parent--child ($k=1$) and grandparent-grandchild correlations ($k=2$).\footnote{We consider the implications of a two-parent structure below.} In the absence of direct multigenerational estimates, researchers sometimes ``iterate'' the parent-child correlation $\beta_{-1}$ to describe the expected rate of persistence across more than two generations. For example, a naive prediction for the grandparent-grandchild correlation $\beta_{-2}$ would be the product of the two underlying parent-child correlations (i.e., $\beta_{-1}^2$ in steady state). 

Alternatively, we may report the slope coefficients from a multigenerational
(child-parent-grandparent) regression such as
\begin{equation}
y_{it}=\beta_{p}y_{it-1}+\beta_{gp}y_{it-2}+\varepsilon_{it},\label{eq:child-parent-gp-regression}
\end{equation}
where $\beta_{gp}$ captures whether grandparent status $y_{it-2}$
has an independent association with child status, even conditional on parent status $y_{it-1}$.\footnote{These are the most common measures of multigenerational mobility, but other measures are in use; see in particular the ``dynastic'' regressions in \citet{Adermonetal2016} and name-based measures in \citet{Clark2014book} and \citet{BaroneMocetti2016}.}

This ``grandparent coefficient'' $\beta_{gp}$ is a useful summary measure of the dynamic properties of the transmission process. In particular, there exists a direct mapping between the coefficients $\beta_{-k}$ from the pairwise regressions in eq. (\ref{eq:pairwise_regression}) and the grandparent coefficient $\beta_{gp}$ from eq. (\ref{eq:child-parent-gp-regression}). Under stationarity, the grandparent coefficient in a regression of child on parent and grandparent
status from the parent's \emph{own} lineage is%
\footnote{See \cite{ECOJ:ECOJ12453}. The stationarity assumption simplifies the presentation but is not necessary for the result (see Appendix \ref{sec:GP-Coeff-Nonstationary}).}
\begin{align}
\beta_{gp} & =\frac{\beta_{-2}-\beta_{-1}^{2}}{1-\beta_{-1}^{2}}\label{eq:Duality_result}
\end{align}
and therefore is positive if and only if $\beta_{-2}>\beta_{-1}^{2}$,
that is, if intergenerational mobility declines at less than the geometric
rate.\footnote{Similarly, the grandparent coefficient on a grandparent from the \emph{other} lineage equals $\beta_{gp}^{\prime} = (\beta_{-2} - \beta_{-1}\beta_{-1}^{\prime}) / (1 - (\beta_{-1}^{\prime})^2)$, where $\beta_{-1}^{\prime}$ is the correlation between a person and their parent-in-law, i.e., $\beta_{-1}^{\prime} = Cov(y_{it-1}^{m}, y_{it-2}^{p,m}) / Var(y_{it-2}^{p,m})$, where $p,m$ denote the paternal or maternal lines.} We label this condition ``excess persistence''. With excess persistence, a naive extrapolation from standard parent-child correlations would understate the persistence of status differences across multiple generations (i.e., overstate mobility in the long run). 

\begin{figure}[t!]
    \centering
    \caption{Multigenerational Estimates for Different Regions}
    \label{fig:model-example1}
        \centering
        \includegraphics[width=8cm]{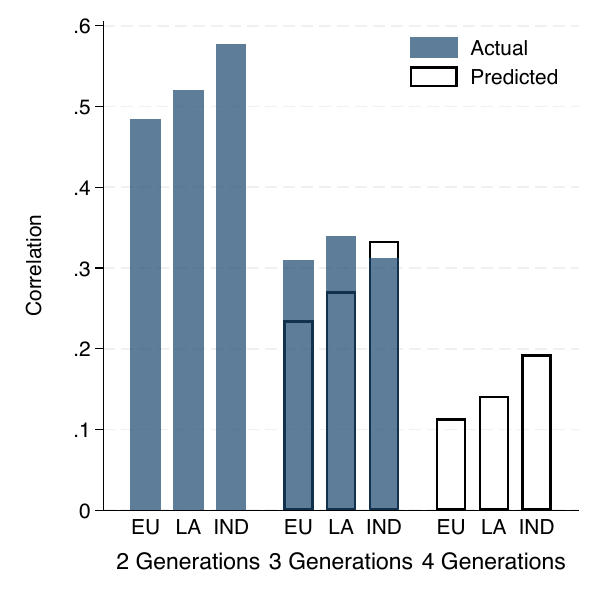}
\begin{minipage}{1.0\textwidth} \setstretch{1}
\bigskip
{\footnotesize \textbf{Note}: The figure shows multigenerational correlations for three regions. The solid bars represent estimates for the EU-28 (from \citealt{colagrossi2020like}), six Latin American countries (\citealt{GallegosCelhay2016}), and Indonesia (IFLS, own estimates). The hollow bars show naive extrapolations from the parent-child correlations (see text).}
\end{minipage}
\end{figure}

Figure \ref{fig:model-example1} compares estimates of the parent-child correlation in education $\beta_{-1}$ and grandparent-child correlation $\beta_{-2}$ for a pooled sample of EU-28 countries (\citealt{colagrossi2020like}), six Latin American countries (\citealt{GallegosCelhay2016}), and Indonesia (based on the Indonesian Family Life Survey, as studied below).\footnote{We report the average of the correlation between generations 1 and 2 and the correlation between generations 2 and 3. Parental education is the average education of both parents.} Although their empirical specifications are not fully comparable, the estimates align with the conventional wisdom that parent-child correlations are higher in low- than high-income countries. Indonesia has the highest parent-child correlation, implying the lowest level of mobility. If we were to extrapolate naively from these parent-child correlations ($\beta_{-1}^k$, hollow bars), we would expect higher multigenerational mobility in Europe compared to the other regions. Indeed, because differences in $\beta_{-1}$ magnify with each additional generation $k$, they become more pronounced in relative terms. 

However, the actual grandparent-child correlations (solid bars) deviate substantially from these naive predictions. For the European and Latin American regions, the correlations between three generations are larger than a naive extrapolation from the parent-child evidence would suggest (``excess persistence", i.e. the coefficient $\beta_{gp}$ from eq. (\ref{eq:child-parent-gp-regression}) is positive). In contrast, for Indonesia the relationship is the opposite, with the actual grandparent-child correlation being below its predicted value. As a consequence, the ranking of the different regions changes; while Indonesia has the lowest mobility according to conventional parent-child measures, its multigenerational mobility is not particularly low. In the next section we discuss why the relation between inter- and multigenerational correlations might vary between more and less developed countries.

\section{Theories of multigenerational transmission} \label{sec:model}

This section provides a theoretical framework for interpreting multigenerational transmission patterns. Our key argument is that these patterns, in particular the relation between short and long-run mobility, will be systematically different in developed and developing countries. One corollary is that our understanding of social mobility in the developing world is still quite limited; while it is well-documented that
\emph{inter}generational (parent-child) mobility is comparatively low (cf. Figure \ref{fig:model-example1}), it does not
necessarily follow that multigenerational mobility is also low in
developing countries.  

We conduct our analysis in three steps. In Section \ref{sec:Model_main}, we derive
implications for long-run mobility -- in particular, the sign of
$\beta_{gp}$ in eq. (\ref{eq:child-parent-gp-regression}) --
from standard models of intergenerational transmission. In Section
\ref{sec:Model_developing}, we explain why these standard models imply different patterns of multigenerational transmission in developing countries. After probing these implications empirically, we then extend our model in Section \ref{sec:mating} to include assortative matching, showing that marital customs are another factor why mobility patterns are bound to differ in developing countries from those observed in developed countries. 

\subsection{Multigenerational transmission in standard models}\label{sec:Model_main}

\textbf{A simplified Becker-Tomes model.} We start with standard models of intergenerational transmission with a one-parent structure, such as the classic Becker-Tomes model of intergenerational transmission (\citealt{becker1979equilibrium}, \citealt{becker1986human}). One frequently cited implication from this model is that $\beta_{gp}$ in eq. (\ref{eq:child-parent-gp-regression}) should be negative. To see this, consider a popular variant of their framework as considered by \citet{solon2004model} or \citet{Piraino:2021},\footnote{While Becker and Tomes considered a behavioral model of parental investments, this behavioral model implies the type of ``mechanical'' transmission equations on which we focus here; see \cite{Goldberger1989}, \cite{solon2004model}, and \cite{Piraino:2021}.} 
\begin{align}
y_{it} & =\rho h_{it}\label{eq:Solon_y}\\
h_{it} & =\theta y_{it-1} + e_{it}\label{eq:Solon_h}\\
e_{it} & =\lambda e_{it-1}+v_{it}\label{eq:Solon_e}
\end{align}
where $y_{it-1}$ and $y_{it}$ are parent and child income in family $i$, $h_{it}$ is child ``human capital'',  $e_{it-1}$ and $e_{it}$ are parent and child ``endowments'' (an umbrella term for a wide set of skills, preferences and other factors that might affect human capital accumulation), and $v_{it}$ is a white noise error term. The parameter $\lambda$ therefore reflects the ``transferability'' of endowments across generations, while $\theta$ represents a ``direct'' effect of parental income on child human capital, possibly due to parental investments in the education of their child in the context of imperfect capital markets (\citealt{becker1986human}, {\citealt{Piraino:2021}). As in this model child income is a deterministic function of child human capital, we can simplify by substituting eq. (\ref{eq:Solon_h}) into (\ref{eq:Solon_y}) and defining $\gamma=\rho\theta$. 

To link this model to the estimating eq. (\ref{eq:child-parent-gp-regression}), consider the difference between (\ref{eq:Solon_y}) and $\lambda$ times its lag,
\[
\begin{array}{crcl}
 & y_{it}-\lambda y_{it-1} & = & \gamma y_{it-1}+\rho e_{it}-\gamma\lambda y_{it-2}-\rho\lambda e_{it-1}\\
\Rightarrow & y_{it} & = & (\gamma+\lambda)y_{it-1}-\gamma\lambda y_{it-2}+\rho v_{it}
\end{array}
\]
which implies that OLS estimation of eq. (\ref{eq:child-parent-gp-regression})
would estimate $\beta_{gp}=-\gamma\lambda$, which is \emph{negative}
(for $\gamma>0$ and $\lambda>0$). This model therefore suggests
that socio-economic inequalities are \emph{less} persistent than a
naive extrapolation of the available parent-child correlations $\beta_{-1}$
would suggest. \citet{Solon201413} describes the intuition for this result: If the
parent does not earn more despite the advantages of higher income in the grandparent
generation, the parent must have poor endowments; and those poor endowments
might then be passed on to the child. 

\textbf{The latent factor model.} The implication $\beta_{gp}<0$, as already emphasized by Becker and Tomes, has long been controversial; \citet{Goldberger1989} calls it an ``artifact''. As it turns out, it is also counterfactual: recent estimates of $\beta_{gp}$ from developed countries are consistently positive \citep{AndersonSheppardMonden2017}. To motivate multigenerational studies, researchers have therefore considered alternative models, such as the \emph{latent factor model} (\citealt{Clark2014book}, \citealt{ECOJ:ECOJ12453}) given by\footnote{The latent factor model is not the only potential framework to rationalize $\beta_{gp}>0$. In particular, grandparents might have a direct influence on the status of their children that is independent from the status of the parents (\citealt{Mare2011}); and these influences could be greater in developing countries, where multigenerational co-residence arrangements are more common. The evidence on this question has been mixed; \citet{ZengXie2014} find that the education of co-resident grandparents in a sample from China is more predictive than the education of absent grandparents. However, a meta study by \citet{AndersonSheppardMonden2017} suggests that the association of child and grandparent education does not vary systematically with contact. Our focus therefore remains on Markovian (i.e., parent-child) transmission models. \cite{StuhlerMultigenerationalHB} reviews alternative models, including higher-order Markov processes and transmission models with ``multiplicity'', in which intergenerational persistence differs for different types of endowments.}
\begin{align}
y_{it} & =\rho e_{it}+u_{it}\label{eq:LF-y}\\
e_{it} & =\lambda e_{it-1}+v_{it}\label{eq:LF-e}
\end{align}
where $u_{it}$ is a white-noise error term, and the other variables are defined as above. If for simplicity we normalize the steady-state variances of both $e$ and $y$ to one,\footnote{This normalization simply involves rescaling the model's parameters and, as such, entails no loss of generality.} we have
\begin{align}
\beta_{-1}\equiv\frac{Cov(y_{it,}y_{it-1})}{Var(y_{it-1})} & =Cov(\rho e_{it}+u_{it},\rho e_{it-1}+u_{it-1}) \nonumber \\
 & =\rho^{2}\lambda \label{eq:LFM-beta1} \\
\beta_{-2}\equiv\frac{Cov(y_{it,}y_{it-2})}{Var(y_{it-2})} & =\rho^{2}\lambda^{2}. \label{eq:LFM-beta2}
\end{align}
As long as $\rho$ is less than one -- which, given the normalizations, is equivalent to assuming  $Var(u)>0$ -- the grandparent-child correlation exceeds the square of the parent-child correlation
in this model (i.e., $\beta_{-2}>\left(\beta_{-1}\right)^{2}$). Given eq. (\ref{eq:Duality_result}), this in turn implies
that the coefficient $\beta_{gp}$ in (\ref{eq:child-parent-gp-regression}) is necessarily \emph{positive} ($\beta_{gp}=\frac{\rho^{2}\lambda^{2}-\rho^{4}\lambda^{2}}{1-\rho^{4}\lambda^{2}}>0$), and socio-economic inequalities are \emph{more} persistent than a naive extrapolation of the available parent-child correlations $\beta_{-1}$
suggests. The intuition for this result is that, in this model, parent-child correlations are attenuated as $y$ is only an imperfect proxy for the ``true'' endowments of a person.

Why do the two models yield opposing implications on multigenerational persistence? The latent factor model abstracts from the ``direct'' income effect ($\gamma=0$) that generates the negative sign of $\beta_{gp}$ in the Becker-Tomes model. Conversely, the simplified variant of the Becker-Tomes model described above imposes a deterministic relation between endowments $e$ and income $y$ (i.e., $Var(u)=0$), thus abstracting from the imperfect link between endowments and status that underlies the positive sign of $\beta_{gp}$ in the latent factor model. 

\textbf{The Becker-Tomes model.} Perhaps less well known is that the \emph{original} Becker-Tomes model has ambiguous  implications for the sign of $\beta_{gp}$; it allows for a stochastic component in the relation between endowments
and income, 
\begin{equation}
y_{it}=\gamma y_{it-1}+\rho e_{it}+u_{it}\label{eq:BT-y}
\end{equation}
and therefore nests both the simplified Becker-Tomes model in eqs. (\ref{eq:Solon_y})-(\ref{eq:Solon_h}) and the latent factor model in eqs. (\ref{eq:LF-y})-(\ref{eq:LF-e}). In this model, regression (\ref{eq:child-parent-gp-regression}) does not yield $\beta_{gp}=-\gamma\lambda$, unless the error term $u$ has zero variance.\footnote{Specifically, the model implies $y_{it} =(\gamma+\lambda)y_{it-1}-\gamma\lambda y_{it-2}+\rho v_{it}+u_{it}-\lambda u_{it-1}$, and because $u_{it-1}$ is correlated with $y_{it-1}$, OLS estimation of eq. (\ref{eq:child-parent-gp-regression}) will not yield an unbiased coefficient $\beta_{gp}=-\gamma\lambda$. Instead, estimates of $\beta_{p}$ will be downward biased and estimates of $\beta_{gp}$ upward biased. See \citet{Lindahletal2013_0002} for an IV approach to address this endogeneity using four generations of data.} Instead, if we normalize the variances of $e$ and $y$ to one, the coefficient can be expressed as (see Appendix \ref{sec:Proof_BT_GP})
\begin{equation}
\beta_{gp}=\frac{\frac{\rho^{2}\lambda}{1-\gamma\lambda}\left(\lambda-\gamma-\frac{\rho^{2}\lambda}{1-\gamma\lambda}\right)}{1-\left(\gamma+\frac{\rho^{2}\lambda}{1-\gamma\lambda}\right)^{2}}\label{eq:BT-original-betagp}
\end{equation}
where the denominator is positive, and the sign of the numerator is ambiguous and depends on the relative sizes of $\gamma$ and $\lambda$. More specifically, since the parent--child correlation in this model equals $\beta_{-1}=\gamma+\frac{\rho^{2}\lambda}{1-\gamma\lambda}$, eq. (\ref{eq:BT-original-betagp}) simplifies to $\beta_{gp}=(\beta_{-1}-\gamma)(\lambda-\beta_{-1})/(1-\beta_{-1}^{2})$ (see Appendix \ref{sec:Proof_BT_GP}), the grandparent coefficient is positive if and only if endowments are more persistent than observed status ($\lambda>\beta_{-1}$). If parental status $y_{it-1}$ has a comparatively large direct effect on child status, that is $\gamma>\lambda$, the coefficient $\beta_{gp}$ will be negative, as in the simplified Becker-Tomes model.\footnote{Since we standardized the variance of $y$, the variance of $u$ is implicitly determined by the other parameters. As a result, eq. (\ref{eq:BT-original-betagp}) does not directly reveal how the variance of $u$ influences the coefficient $\beta_{gp}$. For the non-standardized expression, see eq. (\ref{eq:beta_gp_BT_general}) in the Appendix.} Notably, this is the case Becker and Tomes had in mind, consistent with their view that
intergenerational persistence is primarily due to parental investments in the human capital of their children.\footnote{See footnote 12 in \cite{becker1979equilibrium}. %
\citet{Thomas2000429} notes that ``\emph{Becker and Tomes
(1986) interpret the high level of mobility that they observe in the US primarily in the liberal right-wing way (ability is moderately heritable and markets are highly efficient).}''} 

The more general takeaway however is that the sign and size of $\beta_{gp}$ depend on the \textit{relative} importance of different transmission mechanisms. Figure \ref{fig:model-example2} illustrates this point by plotting the implied multigenerational correlations from two distinct parametrizations of the Becker-Tomes model, one assuming $\gamma=0$ and implying $\beta_{gp}>0$ (blue line) and the other assuming $\gamma>0$ and implying $\beta_{gp}<0$
(red line). While intergenerational mobility is lower in the ``red'' model with direct income effects, multigenerational mobility is lower in the ``blue'' model. The ranking of countries in terms of \textit{inter}generational mobility may therefore tell us little about their ranking in terms of long-run mobility across multiple generations.

\begin{figure}[H]
    \centering
    \caption{Two Simulated Multigenerational Processes}
    \label{fig:model-example2}
        \centering
        \includegraphics[width=8cm]{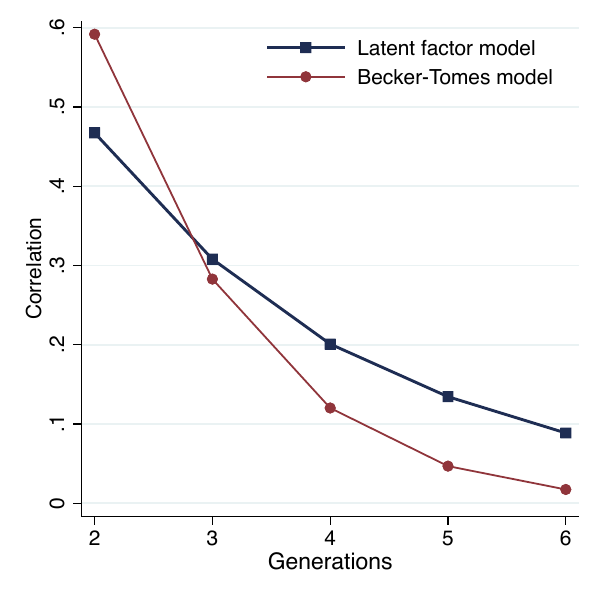}
\begin{minipage}{1.0\textwidth} \setstretch{1}
\bigskip
{\footnotesize \textbf{Note}: The figure shows multigenerational correlations for two simulated processes. The first process is based on the latent factor model in eqs.(\ref{eq:LF-y})-(\ref{eq:LF-e}) with parameters corresponding to the average estimates from \citet{colagrossi2020like}, $\lambda=0.66$ and $\rho=0.84$ (implying $\beta_{gp}=0.11$). The second process is based on the Becker-Tomes model in eq. (\ref{eq:BT-y}) and parameter values $\gamma=0.33$, $\lambda=0.33$ and $\rho=0.84$ (implying $\beta_{gp}=-0.11$).}
\end{minipage}
\end{figure}

\subsection{The role of financial constraints in multigenerational transmission}\label{sec:Model_developing}

As multigenerational patterns depend on the relative importance of different transmission mechanisms, those patterns are bound to differ between developing and developed countries. It is already well-established that developing countries tend to have comparatively low intergenerational mobility; for example, a recent World Bank report by \citet{narayan2018fair} notes that ``\emph{all 15 economies that rank in the bottom 10 percent by relative IGM are developing economies}''; and similar conclusions are found in earlier studies (\citealt{Hertzetal2008}, and \citealt{brunori2013inequality}). 

Less obvious is \emph{why} parent-child mobility is lower in developing countries.\footnote{A related question is how mobility varies with economic growth; see \cite{MaozMoav1999} for a theoretical framework and \cite{Neidhofer2024} for recent empirical evidence.} In the Becker-Tomes model, $\beta_{-1}$ increases in the parameters $\gamma$, $\lambda$ or $\rho$, each of which represent distinct transmission mechanisms. However, the literature generally points to the ``direct'' income effects represented by $\gamma$ as a key mechanism for why $\beta_{-1}$ is larger in developing countries. Specifically, \citet{Piraino:2021} and \citet{mogstad2021family} emphasize the role of liquidity and credit constraints, which are more salient for parental investments in child education in developing countries than in developed countries (\citealt{AttanasioKaufmann2009}, \citealt{Solis2017}).

The hypothesis that the direct income effects represented by $\gamma$ tend to be larger in developing countries has important implications for multigenerational transmission. In Appendix \ref{sec:Proof_BT_GP}, we show that in the Becker-Tomes model the derivative of the grandparent coefficient $\beta_{gp}$ with respect to $\gamma$ is negative ($\frac{\partial\beta_{gp}}{\partial\gamma}<0$). Ceteris paribus, the coefficient $\beta_{gp}$ should therefore be more negative in developing countries in which liquidity and credit constraints are more binding.\footnote{From the ``duality'' in eq. (\ref{eq:Duality_result}), it follows that the ratio between multi- and intergenerational correlations $\beta_{-k}/\beta_{-1}$ would be comparatively low.} Credit constraints would therefore lead not only to high parent-child correlations (as is well-known), but also a $\beta_{gp}$ coefficient that is less positive or even negative (a novel implication). 

Thus, even if parent-child mobility is low in developing countries, \emph{long-run} mobility may not necessarily be low (e.g., as the ``red'' Country in Figure \ref{fig:model-example2}). The intuition for this result is that while credit constraints have a direct effect on educational attainment (increasing parent-child correlations), they do not necessarily affect the transmission of endowments from one parent to the next, and hence have less severe implications for a family's prospects in the long run. In our empirical analysis, we will probe this hypothesis by exploiting information about household educational expenditures as well as geographical variation in credit constraints driven by the 1997 Asian financial crisis. 

Of course, the Becker-Tomes model leaves out many other important mechanisms. Apart from financial and credit constraints, \cite{Piraino:2021} identifies two other important factors that may reduce mobility in the developing world: labor market segmentation, in particular between formal and informal sectors, and informational frictions, on the labor market or in terms of parental beliefs about the returns to educational investments. In Section \ref{sec:mating}, we highlight marital customs and assortative mating as another distinct factor shaping multigenerational dynamics in developing countries.

\section{Data and empirical results} \label{sec:data}
\subsection{Data}
Our main data source is the Indonesian Family Life Survey (IFLS), a nationally representative socioeconomic and health survey covering approximately 83\% of the Indonesian population living in 13 of the nation's 26 provinces \citep{frankenberg1995}.\footnote{The provinces surveyed include North Sumatra, West Sumatra, South Sumatra, and Lampung on the Sumatra island, all five provinces in Java (DKI Jakarta, West Java, Central Java, DI Yogyakarta, and East Java), and four provinces from the remaining major island groups (Bali, West Nusa Tenggara, South Kalimantan, and South Sulawesi). The IFLS surveys individuals also upon movements to other provinces, so in our final sample we have children born in all Indonesian provinces, although provinces not included in the original sampling strategy are clearly less represented.} We pool five waves of the IFLS (1993, 1997, 2000, 2007, and 2014) to form a panel dataset. Since our objective is to explore the transmission of education across three generations, we conduct our analysis using only individuals for whom we have information on their complete education and that of their ancestors, totalling 8,278 individuals. 
Our sample of individuals from three generations is then defined as follows: (1) third-generation individuals, \textit{G1}, were born between 1975 and 1988; (2) their parents, \textit{G2}, who are household heads in the first IFLS wave (the median year of birth is 1952 for fathers and 1957 for mothers); (3) the grandparents, \textit{G3}, of G1 and thus parents of G2, (the median year of birth is 1913 and 1920 for paternal and maternal grandfathers, respectively, and 1922 and 1927 for grandmothers).

Our analysis sample requires complete education information for parents and grandparents, raising the concern that selective attrition, particularly related to migration, could affect representativeness. Appendix Table \ref{tab:select} shows that parent–child education correlations are very similar across samples that differ in the extent of missing ancestral information, indicating little sensitivity to the restriction. Appendix Table \ref{tab:move} further examines selection into the restricted sample and finds no significant association with migration status or other observable characteristics. Taken together, these results suggest that selection into our estimation sample is unlikely to drive our main findings. 

Our outcome of interest is education, which we measure as the highest level of schooling obtained by age 25, by which formal education is typically completed.\footnote{More details on the education measure and steps for validation with the census data are available in Appendix \ref{sec:val}.} In the IFLS, individuals are asked about the different levels of schools they have completed. We translate this information into years of education according to the time it normally takes to complete a particular level. If an individual takes more years than typically needed to complete one specific school level, we still assign to that person the years of education which typically correspond to that school level. As such, our measure of education should be interpreted as the equivalent number of years required to complete one specific school level.

Educational attainment in the IFLS is constructed from multiple survey questions and, in some cases, reported by different household members, which can generate inconsistencies.\footnote{These discrepancies typically arise when both the head of the household and individual household members provide education information. For example, an individual may report having completed secondary education, while the head of household reports that the same individual attained only primary education.} We resolve these inconsistencies in two steps. First, within each survey wave, some individuals appear in the education files with multiple records corresponding to different schooling levels. In these cases, we assign the highest completed education level recorded. Second, across waves, when the same individual reports different education levels in different survey rounds, we assign the modal reported education level among waves in which the individual is observed at age 25 or above.

Importantly for our multigenerational analysis, information on grandparents’ education is reported by the household head for both their own parents and their parents-in-law, including when these individuals are deceased or reside outside the household. This reporting structure may introduce measurement error in grandparental education, particularly for in-laws. To the extent that this attenuates multigenerational correlations, it is relevant for interpreting the additional persistence arising from assortative mating analyzed in Section \ref{sec:mating}.

 In our analysis, we also use information from the IFLS on provinces and district ({\it kabupaten}) of birth and residence, migration, marital traditions, education expenditures and earnings. Similar to provinces, for kabupaten individuals are followed even if they moved from the kabupaten where they were originally surveyed, and our sample includes individuals born in 238 distinct kabupaten. Section \ref{sec:varinfo} in the Appendix provides more details on the construction of these variables. 

\subsection{Descriptive statistics\label{sec:descriptives}}

Table~\ref{tab:descriptives} reports descriptive statistics by generation. The average years of education increased dramatically over the three generations. The increase is more pronounced for women than men and reflects a significant decrease in the proportion of individuals with zero education across generations.\footnote{While more than 70\% of the grandmothers and about 60\% of the grandfathers have zero education, only around 19\% of mothers and 11\% of fathers (in the G2 generation) and almost none in the G1 generation have no schooling (see Figure~\ref{fig:distr_yoe_all}).\label{footnote:zero_edu}} More specifically, differences in mean schooling across generations are partly due to a concentration around zero for the grandparent (G3) generation, more dropouts during primary school and a concentration around six years of schooling (equivalent to primary school completion) for the parent generation (G2), and a concentration around nine years of schooling (completed middle school) and 12 years (completed high school) for the child generation (G1). 

 \begin{table}[t]
       \centering
 \singlespacing
\caption{Descriptive Statistics for the Restricted IFLS Sample
}\label{tab:descriptives}
 \renewcommand{\arraystretch}{0.8} 
       \begin{tabular}{lcccc}
               \toprule
               &G1: Men   & G1: Women & G2: Father & G2: Mother \\
     \midrule
               Median Year of Birth & 1982  & 1982  & 1952  & 1957 \\
               Years of Education   & 9.66     &  9.74      &  6.09   & 4.77   \\
                       \% Migrant & 3.6   & 3.3   & 4.6   & 5.2 \\
               \multicolumn{1}{l}{\% Family Selects the Spouse} & -     & -     & 19.1  & 27.9 \\
               \multicolumn{1}{l}{Age at Childbirth} & -     & -     & 31.8  & 26.1 \\
               \multicolumn{1}{l}{Earnings} & -     & -     & 27,636 & 23,513 \\
 
               Observations & 4,262 & 4,016 & 8,278 & 8,278 \\
\midrule
Lineage                & \multicolumn{2}{c}{Paternal} & \multicolumn{2}{c}{Maternal} \\
G3             & Grandfather & Grandmother & Grandfather & Grandmother \\
\midrule
               Median Year of Birth & 1913  & 1922  & 1920  & 1927 \\
       Years of Education   & 2.30       &  1.34      &  2.59     & 1.48   \\
 
\multicolumn{1}{l}{Age at Childbirth} & 39.4  & 30.8  & 38.1  & 30.5 \\
 Observations & 8,278 & 8,278 & 8,278 & 8,278 \\
               \bottomrule
       \end{tabular}%
\vspace{2mm}
\begin{minipage}{0.94\textwidth} \setstretch{1}
{\footnotesize {Note: The table reports the mean of variables used in the empirical analysis. An individual is defined as a migrant if he/she has changed the province of residence between birth and age 12. Earnings are expressed in thousands of Rupiah. Only some individuals born between 1975 and 1988 report earnings and occupation by 2014, hence we do not report the statistics for this generation (G1). For G2, age at child birth is measured as the average for all children within a household. For G3, age at child birth is measured at the time of birth of G2. We report the averages of this variable without outliers.}}
\end{minipage} 
       
\end{table}%

In terms of marital traditions, the practice of the family selecting the spouse is more common among women than men. About 28\% of the mothers (women in the second generation) in our estimation sample report that their parents selected the spouse. %
Provinces where more than 40\% of the women declare that their spouse has been selected by their parents include South Sulawesi (75\% of the women declare that their husband has been selected by their parents), South Kalimantan (45\%), East Java (44\%), Central Java (43\%), West Sumatra (42\%).

\subsection{Intergenerational correlations} \label{sec:correlations}

We begin the analysis by estimating the intergenerational transmission of education in Indonesia using the restricted sample of 8,278 grandchildren (G1), for whom we have educational outcomes for both of the parents (G2) and all four of the grandparents (G3). %
We first report the intergenerational correlation coefficient for the various pairs of G3 (grandparents), G2 (parents), and G1 (children), by lineage. Table~\ref{tab:igvsmg} shows that the G1-G2 correlation coefficient using the average years of schooling for parents is 0.516 while the G2-G3 correlation is 0.638. These estimates are in line with earlier work, showing that intergenerational mobility in Indonesia is relatively low compared to more developed countries (\citealt{Hertzetal2008}, \citealt{VanderWeideJDE24}).\footnote{While we focus on linear summary measures, \cite{Ahsanetal2024JEBO} document important non-linear patterns in Indonesia. In particular, while the conditional expectation function for children's schooling given parental schooling is linear in urban areas, it is convex in rural Indonesia.}

 \begin{table}[t]
 \singlespacing
    \centering
            \caption{Estimated Intergenerational Correlation}\label{tab:igvsmg}
    \begin{tabular}{llllllll}
    \toprule
        \multicolumn{2}{c}{Lineage} & ~ & \multicolumn{3}{c}{Observed} & ~ & \multicolumn{1}{c}{Predicted} \\
        \cline{1-2}\cline{4-6}\cline{8-8}  
        G2 & G3 & ~ & G1-G2 & G2-G3 & \multicolumn{1}{l}{G1-G3} & ~ & \multicolumn{1}{l}{G1-G2xG2-G3} \\ 
 
        \hline
                Both (avg) & All (avg) & ~ & 0.516 & 0.638 & 0.312 (0.014) & ~ & 0.329 (0.010) \\ 
        ~ & ~ & ~ & ~ & ~ & ~ & ~ & \\ 

        Father & Paternal (avg) & ~ & 0.486 & 0.507 & 0.247 (0.015) & ~ & 0.247 (0.009) \\ 
        Father & Maternal (avg) & ~ & 0.486 & 0.491& 0.292 (0.014) & ~ & 0.239 (0.010) \\ 
        ~ & ~ & ~ & ~ & ~ & ~ & ~ & \\ 
        Mother & Paternal (avg) & ~ & 0.464 & 0.469&  0.247 (0.015) & ~ & 0.218 (0.010) \\ 
        Mother & Maternal (avg) & ~ & 0.464 & 0.567 & 0.292 (0.014) & ~ & 0.263 (0.010) \\ 
        \bottomrule
    \end{tabular}

    \vspace{2mm}

\begin{minipage}{0.88\linewidth}
{\footnotesize Note: G1 denotes the child generation; G2 denotes the parent; G3 denotes the grandparents. Total number of observations is 8,278. Correlations are reported for G1-G2 and G2-G3. Regression coefficients with robust standard errors clustered at the family level in parentheses are reported for G1-G3.}
\end{minipage}

\end{table}

\subsection{Multigenerational correlations} \label{sec:mcorrelations}

We next estimate the transmission of education across three generations. As shown in Table~\ref{tab:igvsmg}, the G1-G3 coefficients vary between 0.25 and 0.31, depending on specification. To understand whether these three-generation estimates suggest more or less persistence than what one might have expected based on the available two-generation estimates, we also compute the predicted three-generation correlation as would be implied by the product of the corresponding parent-child correlations (see Section \ref{sec:measurement}). Those predicted correlations are not so different from the actual three-generation correlation: they are slightly smaller when considering individual lineages (rows 1-4), but slightly larger if averaging across all members of a given generation (last row). As already illustrated in Figure \ref{fig:model-example1}, this pattern contrasts with the pattern found for developed countries, in which multigenerational persistence is generally higher than implied by parent-child correlations (\citealt{AndersonSheppardMonden2017}). Therefore, while \textit{inter}generational correlations are comparatively high in Indonesia, \textit{multi}generational correlations are not.

Table~\ref{tab:maintable} reports estimates from three-generation regressions based on equation (2), using a common restricted sample of 8,278 individuals across all specifications. Column (1) presents baseline results without fixed effects and yields a negative and statistically significant coefficient on grandparents’ education. Column (2) adds ethnicity fixed effects, while column (3) instead includes \textit{kabupaten} fixed effects. Although the grandparent coefficient becomes less precisely estimated when conditioning on \textit{kabupaten} fixed effects, its magnitude remains negative. Column (4) controls for indicators capturing whether parents or grandparents have zero years of education, and the estimated grandparent coefficient remains negative and statistically significant.\footnote{As discussed in footnote \ref{footnote:zero_edu}, among grandparents it is common not to have any education, with 70\% of grandmothers and 60\% of grandfathers reporting zero years of schooling. However, because here we use the {\it average} grandparental education as a control, the proportion of grandchildren who report zero average grandparental education is smaller, around 43\%.} Finally, we investigate potential misspecifications of the estimated multigenerational equation. In column (5) we use a linear probability model where the binary outcome indicates whether the G1 individual completed high school; the estimated grandparent coefficient is negative, comparably large in magnitude and statistically significant. In Appendix Table \ref{tab:nonlinearities} we control for parents and grandparents having no education and further include quadratic terms for parental and grandparental education (columns 2–4) or a dummy for whether grandparental education equals or exceeds parental attainment (column 5). All these specifications confirm that more educated grandparents have {\it less} educated grandchildren once we condition on parental education.\footnote{The parameters estimated in Table \ref{tab:nonlinearities} become less precisely estimated once quadratic terms are included for both parents and grandparents. Overall, they confirm a negative coefficient for grandparents when they have high average education. This same result is confirmed when investigating the data graphically; in Figure \ref{fig:multigen_plot} we show the relationship between children's and grandparents' years of education, separately for parents with different education levels. Such a relationship is clearly non-positive: it is generally flat at low levels of grandparental education and turns negative at higher levels of education, especially for parents whose higher educational attainment is relatively low (primary school). Finally, to check whether the finding of convex {\it inter}generational conditional expectation functions for children’ in rural regions documented in \cite{Ahsanetal2024JEBO} has distinct multigenerational implications, we compare the multigenerational transmission patterns in rural and urban areas in our sample. Table \ref{tab:rural} shows that the grandparent coefficient is more positive in rural regions, suggesting that non-linearities in the intergenerational relationship may also map into distinct multigenerational patterns.}

Overall, we find a consistently negative association between grandparental education and grandchildren’s educational attainment conditional on parental education. This result is consistent with the simplified Becker and Tomes model, which predicts that, conditional on parental education, higher grandparental education reduces grandchildren’s educational outcomes (see Section~\ref{sec:Model_main}). However, our estimates contrast with the positive or null grandparent effects reported for developed countries (e.g., \citealt{AndersonSheppardMonden2017}; \citealt{narayan2018fair}), Latin America \citep{CelhayGallegosChile2015,GallegosCelhay2016}, and China \citep{ZengXie2014}.\footnote{\cite{ZengXie2014} finds a positive coefficient among co-resident grandparents and no effect for non-co-resident grandparents.} 
\\

\begin{table}[t]
\singlespacing
        \caption{Multigenerational Correlations in the IFLS Sample \label{tab:maintable}}  
			\begin{tabular}{l*{5}{c}}
				\hline
				&\multicolumn{1}{c}{(1)}&\multicolumn{1}{c}{(2)}&\multicolumn{1}{c}{(3)}&\multicolumn{1}{c}{(4)}&\multicolumn{1}{c}{(5)}\\
				Outcome Variable&\multicolumn{4}{c}{Years of Education}&\multicolumn{1}{c}{1[High School]}\\
\hline
Parental Education 		&       0.502\sym{***}&       0.501\sym{***}&        0.413\sym{***}	&       0.484\sym{***}&     0.0656\sym{***}\\
						&    (0.0144)         &    (0.0135)         &    (0.0153)         	 &    (0.0153)        &   (0.00188)         \\ 
Grandparental Education	&     -0.0426\sym{*}  &     -0.0376\sym{*}         &    -0.0175         	&     -0.0539\sym{*}   &    -0.00705\sym{**} \\
						&    (0.0238)         &    (0.0218)         &    (0.0246) 			&    (0.0293)         &   (0.00337)         \\
						
&   &   &   &   &   \\ 
Ethnicity FE$^a$    	& no & yes & no & no & no  \\
Kabupaten FE$^b$  		& no &no & yes  & no & no \\
P and GP with no education FE & no &no & no  & yes & no \\
&   &   &   &   &   \\ 
Observations        &        8,278         &        8,278         &        8,278          &        8,278      &   8,278      \\
R-squared           &       0.267         &       0.350        &       0.355          &       0.268         &       0.229          \\
\hline

\end{tabular}\\
  \vspace{2mm}\centering
\begin{minipage}{0.97\linewidth}
	{\footnotesize  $^a$Ethnicity is only asked to respondents in the 2000, 2007, and 2015 waves. In our sample, 1,049 individuals have missing information on ethnicity. In this specification we included a dummy for missing information on ethnicity, as well as ethnicity dummies, as controls.\\
		$^b$260 observations have missing information about their Kabupaten of residence. In this specification we included a dummy for missing information on Kabupaten, as well as Kabupaten dummies, as controls.\\
		\textbf{Note}:  Multigenerational estimates from eq. (\ref{eq:child-parent-gp-regression}), the dependent variable is the child's years of education (G1) in columns (1)--(4) and a dummy for the completion of high school by the child in column (5). Control variables are the average years of education of the parents (G2) and the average years of education of all four grandparents (G3). Standard errors are clustered at the family level.*** p$<$0.01, ** p$<$0.05, * p$<$0.1.}
\end{minipage}
\end{table}

\section{Evidence on financial constraints and direct income effects} \label{sec:credit} 

In the Becker-Tomes model, $\gamma$ measures the strength of the direct effect of parental income on child human capital. This reduced-form effect may reflect various mechanisms, one of which is the ability of higher-income parents to invest in human capital enhancing inputs, such as educational resources, play materials, high-quality childcare and schooling, as well as extra-curricular experiences.\footnote{Another theory about the link between family income and child outcomes is the family stress channel, which posits that parental income (or lack thereof) has an impact on parental stress and the quality of parent-child interactions, which have an important influence on child human capital.} Its strength thus also depends on parents' ability to borrow: the stronger the credit constraints, the stronger the link between parental income and parental investments that affect child outcomes (\citealt{becker1986human}). Such income effects are thought to be a key mechanism to explain why intergenerational correlations are larger in developing countries. But as shown in Section \ref{sec:model}, conditional on parent-child transmission, credit constraints have the opposite implication for the speed of decay of multigenerational correlations: the higher $\gamma$, the more negative the coefficient $\beta_{gp}$. 

In this section, we provide evidence to test this implication of the Becker-Tomes model regarding the relationship between $\gamma$ and the strength of the grandparent coefficient $\beta_{gp}$. We conduct this analysis not just for the pooled sample, but also by gender. There is a vast literature documenting large gender differences in educational attainment in developing countries.  In India, for example, girls are less likely to be enrolled in private schools relative to boys in the same household (\citealt{Maitraetal2016}). In Indonesia, households that become credit constrained due to crop loss have been shown to cut back school expenditures, particularly so if the child is female rather than male (\citealt{CameronWorswick2001}). In line with the idea that parents favour investments in sons, \citet{ahsan2023public} find that following a general expansion in primary school attainment due to the Inpres program in the 1970s, male students crowded out females in secondary schools, most likely due to gender norms within families. Lower parental propensity to invest in daughters may be driven by a number of factors, including cultural norms that view boys as contributing to old-age security in some contexts (\citealt{Maitraetal2016}), as well as gender differences in the actual or perceived labor market returns to education.\footnote{The latter is supported by empirical studies showing that factors which improve the job prospects for women, such as manufacturing growth in Bangladesh (\citealt{HeathMobarak2015}) or access to jobs in the business process outsourcing industry in India (\citealt{Jensen2012}), result in significantly higher educational attainment for girls.} 

Regardless of the driving factor(s), if parents have a different propensity to invest in their sons and daughters' education, then the link between parental income and child education would likely also differ by gender. However, it is a priori unclear whether we would expect a stronger link for boys or for girls. If parents allocate only a basic amount for daughters  while investing the remainder in sons, the correlation between parental income and educational investment will be stronger for boys. If, on the other hand, parents prioritize investing in sons and invest in daughters only if sufficient financial resources are available, the pattern would be the opposite. 

Descriptive evidence confirming the relevance of the direct effect of parental income channel and of its heterogeneous impact across children's gender is provided in Table \ref{tab:area_share}, where we interact the multigenerational equation with indicators of the educational expenditure share at the province level.\footnote{The IFLS data include information on household expenditure on children's education. For each province, we construct a measure of total educational expenditure relative to aggregate province-level income for households with children. The average value of the share of household income allocated to children's schooling across the 13 provinces originally included in the IFLS is 0.27, with a standard deviation of 0.12. For a more detailed description of the result in Table \ref{tab:area_share}, see Appendix \ref{app:expenditure}.} The grandparental coefficient is more negative in areas where families spend more for the education of children and this negative association is driven by boys (in G1) rather than girls, suggesting that parental resources may be more relevant for sons' education, as families may decide to invest more in male children. However, residential sorting may correlate with unobserved individual characteristics and educational expenditure decisions are potentially endogenous.  As such, we next investigate %
the impact of financial constraints on the multigenerational transmission of education within a more robust causal framework, exploiting an exogenous shock to parental resources%
. While we cannot directly link the exogenous shock to educational expenditures, this causal framework lends further credibility to our hypothesis that financial constraints tend to reduce the size of the grandparent coefficient.

\subsection{The 1997-98 Asian Financial Crisis} \label{sec:shocks}

To isolate the exogenous variation in credit constraints across areas, we leverage geographical variation in the effect of the 1997-98 Asian Financial Crisis on local economic conditions and credit access. The crisis was triggered by the collapse of the Thai baht in 1997, with effects felt in nearby economies soon after (\citealt{Kusnanto2002}).  In January 1998, the Indonesian rupiah had fallen nearly 80\% from its pre-crisis value against the dollar. The crisis led to a 13.7\% economic contraction and an inflation rate as high as 77.6\% in 1998 across the affected countries, with Indonesia, Korea, and Thailand being the three hardest-hit economies (\citealt{imf98}). In Indonesia, it led to a sharp drop in educational attainment, especially for boys (\citealt{Poppele}). 

However, and importantly for our research design, there were vast regional differences in the way that the crisis affected households. Broadly speaking, the provinces of Java and Bali were the most severely hit by the crisis, while the provinces of Sumatera, Sulawesi and Maluku were less affected by or even benefited from the crisis due to elevated prices of their traded products. The provinces in West and East Nusatenggara, Kalimantan and Irian Jaya were somewhat affected, but also suffered from an El Ni\~no drought and forest fires at the same time as the crisis (\citealt{Kusnanto2002}). 

The timing of the crisis and the fact that our data spans cohorts born between 1975 and 1988 mean that we can not only exploit geographical variation in households' exposure to the crisis but also variation across cohorts in a difference-in-differences (DiD) design. Specifically, children who were born in 1978 or before and therefore were 18 or older when the crisis hit were too old for their education to be affected by the crisis.\footnote{In our sample, 99\% completed 16 or fewer years of education (and hence finished by age 22) and 85\% completed fewer than 12 years of education (and hence finished by age 18).} In contrast, children born after 1978 could have seen parental investments in their education affected. 

\subsection{Difference-in-differences framework and measure of economic shock}
To isolate the link between credit constraints on the grandparent coefficient in multigenerational regressions, we need an area-based measure of exposure to the crisis. Our preferred specification uses the proportionate change in DMSP-OLS nighttime lights between 1996 and 1998 in the kabupaten of residence recorded in the 1993 survey  (before the crisis, in order to rule out potential endogenous migratory responses to the crisis). Nighttime light intensity is available annually at very fine spatial resolution from the DMSP-OLS satellite program, which has global coverage from 1992 onward  \citep{noaa_dmsp_ols}. Several papers have validated nighttime lights as a proxy for sub-national economy activity in developing countries, including in Indonesia specifically \citep{henderson2012}, and we provide in \autoref{fig:did_crisis_validation} several pieces of evidence supporting the use of such measure for our purpose.\footnote{\autoref{fig:did_crisis_validation} provides evidence to assess the validity of the kabupaten-level nighttime lights measure as a proxy for local economic shocks in our setting. First, the distribution of changes in nighttime light intensity between 1996 and 1998 exhibits substantial cross-kabupaten variation. The main limitation of nighttime lights is that lights capture formal/industrial economy activity better than agriculture, and very rural districts may be near the floor of the sensor's detection threshold. However, this limitation is less salient in our case, given that the 1997-98 Indonesian crisis hit the urban, formal and industrial sectors hardest. Nevertheless, we show in subfigure (b) that there is variation in the change in nighttime lights across both predominantly urban and predominantly rural kabupaten, though – as expected – the largest declines are concentrated in urban areas. Third, subfigure (c) shows that changes in nighttime lights are positively correlated with province-level changes in GDP over the same period (correlation coefficient = 0.2), while still displaying considerable within-province variation. Finally, subfigure (d) shows that they are also positively correlated with a kabupaten-level Bartik-style predictor of economic shocks, constructed using 1995 employment shares across nine industries and national industry-level GDP changes between 1996 and 1998. Taken together, these patterns support the interpretation of nighttime lights as a meaningful proxy for local economic exposure to the crisis.} In order to interpret the coefficients associated with this measure of crisis, we construct it such that an increase in the variable captures a \emph{worsening} of local economic conditions and standardise it to have a standard deviation of one. 

We operationalise the difference-in-differences approach by estimating the following regression: 
\begin{eqnarray}
    Y_{ick} &=& \beta_{1} EduP_{ick} +  \beta_{2} EduGP_{ick} \nonumber \\
   & +& \beta_{3} EduP_{ick} \times Post_{c} +   \beta_{4} EduGP_{ick} \times Post_{c}  \nonumber \\
     & +& \beta_{5} EduP_{ick} \times Crisis_{k}  +   \beta_{6} EduGP_{ick} \times Crisis_{k} \nonumber \\
    & +&   \beta_{7} EduP_{ick} \times Post_{c} \times Crisis_{k} +  \beta_{8} EduGP_{ick} \times Post_{c} \times Crisis_{k}   \nonumber \\
    & +&    \beta_{9} Post_{c} \times Crisis_{k} + Post_{c} + \delta_{k} + \epsilon_{ick} \label{eq:did}
\end{eqnarray}
where $Y_{ick}$ is the years of education of individual $i$ living in kabupaten $k$ in 1993 and born in year $c$, $EduP_{ick}$ their parents' average education and $EduGP_{ick} $ their grandparents' average education; $Crisis_{k}$ is the kabupaten $k$-level measure of exposure to the crisis defined above;  $\delta_{k}$ are kabupaten fixed effects; and $Post_{c}$ is a dummy indicating whether individual $i$ was born late enough to have their education potentially affected by the crisis (i.e. born in or after 1979). Standard errors are clustered at the karesidenan level, which is a geographical unit in between kabupaten and province.\footnote{Standard errors are clustered at the karesidenan level to account for spatial correlation in local economic shocks. Due to the limited number of provinces (13), province-level clustering would produce downward-biased standard errors. Conversely, clustering at the kabupaten level would fail to account for broader spatial correlation in shocks. The chosen level provides a balance between capturing spatial dependence and ensuring a sufficient number of clusters. \autoref{tab:did_crisis_robust} shows results clustering standard errors at the kabupaten and province levels.}

In the model above, the coefficient of interest is $\beta_{8}$, which indicates how the coefficient on the grandparental education changes among the treated cohorts as they are more exposed to the crisis (and more credit-constrained). Our hypothesis is that $\beta_{8}$ is negative and especially so for groups whose education were more responsive to credit constraints.

In the specification above, the `treatment' group includes all cohorts who were potentially still in school in 1997 (children born in 1979 and later, as opposed to children born between 1975 and 1978). However, while all children younger than 18 at the time of the crisis could have seen their final education affected,  the crisis may have had a differential impact on their final education.  Specifically, we hypothesize that children who were more advanced in their education at the time the crisis hit may have experienced a greater impact of the crisis than children who were younger and who would therefore have had more time to adjust back after the crisis once credit constraints relax. For this reason, after estimating the specification above, we estimate another specification of the model  whereby we distinguish three potentially treated groups of children, defined by the stage of education they were in at the time the crisis hit. Specifically, we group cohorts who were of upper secondary school age (born between 1979 and 1982), cohorts who were of lower secondary school age (born between 1983 and 1985), and cohorts who were of primary school age (born between 1986 and 1988) in 1997. 

\subsection{Results}

Table \ref{tab:did_crisis_main} reports the results. Focusing on Specification 1 (pooling all treated cohorts together), the coefficient $\beta_{8}$ is negative for the full sample, though small and non-significant.  Columns 2 and 3 show results split by gender, revealing  that for boys the coefficient is more negative than it is for girls (for which the coefficient is positive). Nevertheless, in neither of these three samples the coefficient of interest is statistically significant.

\begin{table}[htbp]
\singlespacing
\centering
\caption{Crisis Impact on Multigenerational Links in Education (Difference-in-Differences) \label{tab:did_crisis_main}}
\begin{tabular}{lccc@{\hspace{1em}}ccc}
\hline
  & \multicolumn{3}{c}{Specification 1} & \multicolumn{3}{c}{Specification 2} \\
  & \multicolumn{3}{c}{Pooling treated cohorts} & \multicolumn{3}{c}{Separating treated cohorts} \\
  \cmidrule(lr){2-4}\cmidrule(lr){5-7}  

Sample & All & Males & Females & All & Males & Females \\
 & (1) & (2) & (3) & (4) & (5) & (6) \\

\hline
GP edu & -0.0023 & -0.0075 & 0.0035 & -0.0098 & -0.0221 & 0.0187 \\
 & (0.0493) & (0.0554) & (0.0535) & (0.0515) & (0.0625) & (0.0652) \\
GP edu x crisis & -0.0233 & 0.0186 & -0.0965** & -0.0015 & 0.0689 & -0.1020 \\
 & (0.0516) & (0.0692) & (0.0465) & (0.0492) & (0.0579) & (0.0614) \\
GP edu x Born 79--88 & 0.0031 & 0.0093 & -0.0118 &  &  &  \\
 & (0.0461) & (0.0489) & (0.0662) &  &  &  \\
GP edu x Born 79--82  &  &  &  & 0.0454 & 0.0556 & 0.0100 \\
 &  &  &  & (0.0458) & (0.0612) & (0.0728) \\
GP edu x Born 83--85  &  &  &  & 0.0251 & 0.0266 & -0.0136 \\
 &  &  &  & (0.0564) & (0.0703) & (0.0964) \\
GP edu x Born 86--88 &  &  &  & -0.0139 & 0.0227 & -0.0661 \\
 &  &  &  & (0.0771) & (0.0886) & (0.0895) \\
GP edu x Born 79--88 x crisis & -0.0152 & -0.0471 & 0.0446 &  &  &  \\
 & (0.0458) & (0.0573) & (0.0557) &  &  &  \\
GP edu x Born 79--82 x crisis &  &  &  & -0.0491 & -0.1190** & 0.0360 \\
 &  &  &  & (0.0492) & (0.0452) & (0.0822) \\
GP edu x Born 83--85 x crisis &  &  &  & -0.0637 & -0.1490** & 0.0466 \\
 &  &  &  & (0.0624) & (0.0643) & (0.1030) \\
GP edu x Born 86--88 x crisis &  &  &  & -0.0137 & -0.0753 & 0.0628 \\
 &  &  &  & (0.0543) & (0.0729) & (0.0635) \\
&&&&&&\\
Observations & 7,247 & 3,710 & 3,537 & 7,247 & 3,710 & 3,537 \\
R-squared & 0.3570 & 0.3500 & 0.3950 & 0.3600 & 0.3530 & 0.4000 \\
\hline
\end{tabular}

\vspace{0.5em}
\parbox{\textwidth}{\footnotesize
\textit{Notes:} OLS regression of years of education on average parental education and average grandparental education, fully interacted with binned cohort dummies and the variable ``crisis'', which measures exposure to the crisis as the proportionate change in nighttime lights in the kabupaten of residence in 1993 between 1996 and 1998. This measure is then standardised to have mean 0 and standard deviation 1 in the analysis sample. All specifications also control for kabupaten fixed effects and cluster the standard errors at the Karesidenan level, which is a geographical unit larger than kabupaten but smaller than province. In Specification 1, we bin cohorts born between 1979 and 1988 together as the cohorts whose education could be potentially affected by the crisis and cohort born between 1975 and 1978 as those cohorts who were too old to see their education affected. In Specification 2, we bin the potentially affected cohorts into 3 groups rather than 1 depending on whether they were in upper secondary school (born 79--82), lower secondary school (born 83--85), or primary school (born 86--88) in 1997 when the crisis hit Indonesia. The rest of the coefficient estimates are shown in \autoref{tab:did_crisis_app}. *** $p<0.01$, ** $p<0.05$, * $p<0.1$.
}
\end{table}

\doublespacing

We then turn to our second specification, distinguishing between treated cohorts who were of upper secondary, lower secondary, and primary school age in 1997, in columns 4 to 6. The coefficients of interest remain negative for the sample pooling boys and girls though larger for cohorts who were in lower and upper secondary school at the time of the crisis than for cohorts who were in primary school. Splitting by gender, we see that the coefficients are much more negative and statistically significant at the 5\% for boys who were in lower and upper secondary school when the crisis hit. For girls, all the coefficients are positive but not statistically significant.\footnote{To further validate our hypothesis that the impact of the crisis on the grandparent coefficient was stronger among those who were more advanced in their educational journey and therefore had less time to adjust, columns 4 and 5 of \autoref{tab:did_crisis_robust} replicate Specification 2 for boys, this time using a dummy for completing lower secondary school (at least 9 years) and a dummy for completing upper secondary school (at least 12 years). When the outcome is the likelihood of completing lower secondary school, the impact of the crisis on the grandparent coefficient is significant only for cohorts who were of lower secondary school age at the time of the crisis. In contrast, when the outcome is the likelihood of completing upper secondary school, the impact of the crisis on the grandparent coefficient is significant for the cohort who were of both lower secondary and upper secondary school age at the time of the crisis.}  
 
In \autoref{tab:did_crisis_placebo}, we present the results of a placebo test to support the validity of our strategy. We re-estimate the same specification using only pre-treatment cohorts (born between 1975 and 1978), assigning a placebo treatment whereby cohorts born in 1977 and 1978 are considered potentially affected and cohorts born in 1975 and 1976 are the control group. Given the small sample size, the precision of these placebo estimates is limited. However, for males, the coefficient $\beta_{8}$ is very close to 0, suggesting no systematic pre-trends. For females, the coefficient is negative, suggesting that the positive coefficient reported in \autoref{tab:did_crisis_main} for females may be under-estimated.   

Overall, the empirical analysis presented in this section is consistent with the implications of the Becker-Tomes model about a potential link between financial constraints and the strength of multigenerational transmission. However, this link appears pronounced only for boys, which could reflect that the direct income effect is stronger for boys in the Indonesian context. While it is beyond the scope of this paper to understand why it is the case, this pattern could be consistent with parents having stronger preferences for their son's education over their daughters'.

\section{Assortative mating} \label{sec:mating}

An estimation that omits the education of one parent may lead to inflated multigenerational coefficients, as part of the correlation in education between the omitted parent and the offspring is captured by the education of the grandparents \citep{AndersonSheppardMonden2017}. The size of this omitted variable bias depends on the degree of \textit{assortative mating}, commonly measured as the similarity between spouses in their educational attainment. 

More broadly, assortative mating affects multigenerational patterns. This observation is especially important for cross-country comparisons, as differences in multigenerational estimates may partly reflect differences in the structure and degree of assortative mating, in addition to differences in the intergenerational transmission processes. Cultural norms regarding the process by which spouses match tend to differ significantly between developed and developing countries. Marital customs also vary greatly between developing countries and between ethnic or religious groups. In high-income countries, the spousal search tends to be ``direct'', in that individuals choose their partner themselves. In contrast, in many developing countries, the selection process involves other members of the family -- typically, the parents of the spouses.

In this section, we first illustrate that assortative mating has distinct multigenerational implications, which depend on the structure of the assortative process (Section \ref{sec:Model_ass_mating}). We then show that ``family-based matching'' is associated with larger multigenerational correlations (\ref{subsec:familyAM_evidence}).

\subsection{The role of assortative mating in multigenerational transmission}\label{sec:Model_ass_mating}

To understand the implications of assortative mating – and of different \emph{structures} of the assortative process – consider a version of the latent factor model that incorporates assortative mating. Specifically, assume that a child's endowment is determined by the average of the father's and mother's endowments,
\begin{align}
y_{i,t} & =\rho e_{i,t}+u_{i,t}\label{eq:latent-factor-y}\\
e_{i,t} & =\tilde{\lambda}\bar{e}_{i,t-1}+v_{i,t},\label{eq:latent-factor-e}
\end{align}
with $\bar{e}_{i,t-1}=(e_{i,t-1}^{m}+e_{i,t-1}^{p})/2$, and where the $m$ and $p$ superscripts denote the maternal and paternal lineage. This simple
model is sufficient to derive our key implications, but it can be extended to allow for direct transmission effects, gender-specific transmission pattern, or assortative matching in multiple dimensions (e.g., \citealp{KinshipCorrelationsandIntergenerationalMobility}).

\paragraph*{Direct assortative mating. }

First, consider the implications of ``direct'' assortative mating between spouses based on their latent endowments (normalized to variance one), as represented by the linear projection
\begin{equation}
e_{i,t-1}^{m}=me_{i,t-1}^{p}+w_{i,t-1}\label{eq:latent_direct_AM}
\end{equation}
where $m=Cov(e_{i,t-1}^{m},e_{i,t-1}^{p})$. As in the model without assortative mating, the multigenerational
correlations can still be expressed as
\[
\beta_{-k}=\rho^{2}\lambda^{k}
\]
but $\lambda$ is now defined as $\lambda=Cov(e_{i,t},e_{i,t-1}^{x})=\tilde{\lambda}\left(1+Cov(e_{i,t-1}^{m},e_{i,t-1}^{p})\right)/2=\tilde{\lambda}\left(1+m\right)/2$. Intuitively, the transferability of endowments from one parent to the child as captured by $\lambda$ reflects both the transferability
of the \emph{average} parental endowments ($\tilde{\lambda}$) and the extent of assortative mating between parents ($m$). As $m\leq1$ it follows $\lambda\leq\tilde{\lambda}$, implying that parent-child
correlations are expected to be lower when using the outcome of only one parent to measure parental status, as opposed to the average over both parents.%

Moreover, the size of the grandparent coefficient now depends on which lineages we include in the regression. The grandparent coefficient in a regression of offspring status on parent and grandparent status from the \textit{same} lineage (i.e., father and paternal grandparent from the paternal lineage $p$) still equals (see Appendix \ref{sec:latent_lineages} for details)
\begin{equation}
\beta_{gp}=\frac{\rho^{2}\lambda^{2}-\rho^{4}\lambda^{2}}{1-\rho^{4}\lambda^{2}},\label{eq:betagp_directAM}
\end{equation}
as in the one-parent version of the latent factor model, but with the parameter $\lambda$ now also reflecting the strength of the assortative process.

Conversely, the grandparent coefficient in a regression of offspring status on parent and grandparent status from \textit{different} lineages (i.e., father and maternal grandparent, or mother and paternal grandparent) equals
\begin{equation}
\beta_{gp}^{\prime}=\frac{\rho^{2}\lambda^{2}-m\rho^{4}\lambda^{2}}{1-m^{2}\rho^{4}\lambda^{2}},
\end{equation}
implying $\beta_{gp}^{\prime}>\beta_{gp}$ if assortative mating is
imperfect ($0\leq m<1$). In Appendix \ref{sec:latent_lineages}, we also provide the corresponding expression when including \textit{both} parents. 

We therefore expect the grandparent coefficient to be less positive when including grandparent(s) from the \emph{own} lineage (e.g., including father and paternal grandmother/grandparents), and more positive when including a parent from the \emph{other} lineage
(e.g. father and maternal grandmother/grandparents).\footnote{Note that in the latent factor model considered here, the grandparent coefficient varies depending on which lineages are included in the regression, but always remains non-negative. However, this coefficient can be negative in the Becker-Tomes type models with direct income effects, as considered in Section \ref{sec:Model_main}.} We confirm this hypothesis in Table \ref{tab:lineages}. Column 1 replicates our main estimate from Table \ref{tab:maintable}. In columns 2 and 4, we include the educational outcome of one parent (G2) and the average educational outcomes among grandparents from the \textit{same} lineage (G3);\footnote{As we show in Appendix Table \ref{tab:multi_lineages}, the patterns are similar when considering the education of each grandparent separately.} in columns 3 and 5, we instead include the grandparents from the \textit{other} lineage. 

First, a comparison of column 1 and the other columns illustrates that by excluding one lineage from the estimation of the multigenerational process, we incur an omitted variable bias that results in a more positive grandparent coefficient. This finding is in line with prior evidence (e.g. \citealt{AndersonSheppardMonden2017}, \citealt{ECOJ:ECOJ12453}, \citealt{colagrossi2020like}). Second, the grandparent coefficient tends to be more positive when including parents and grandparents from different lineages (compare columns 2 and 3), in line with our theoretical predictions. Multigenerational and assortative patterns are thus intertwined. As we discuss next, multigenerational correlations also depend on the underlying \textit{structure} of the assortative process.

\begin{table}[t]
\singlespacing
\centering
        \caption{Multigenerational Correlations in Different Lineages} \label{tab:lineages}
\begin{tabular}{l*{5}{c}}
\toprule
    &\multicolumn{5}{c}{Dependent Variable: Child's Education}    \\
 \hline
    {\it Parent}      & Average &\multicolumn{2}{c}{Father}  &\multicolumn{2}{c}{Mother}      \\
    {\it Grandparents}    & Average  & Paternal   & Maternal   & Maternal    & Paternal       \\
             & (1)   & (2)   & (3)   & (4)   & (5)   \\
    \midrule
    Parental Education  & 0.502\sym{***} & 0.404\sym{***} & 0.376\sym{***} & 0.399\sym{***}  & 0.404\sym{***}   \\ 
                       & (0.0144)        & (0.0118)           & (0.0118)       & (0.0140)        & (0.0129)   \\ 
         &   &   &   &   &   \\ 
    Grandparental Education   &  -0.0426\sym{*}  & 0.002     & 0.0898\sym{***} & 0.0539\sym{***} & 0.0505\sym{***}      \\ 
                               & (0.0238)          & (0.0197)   & (0.0180)       & (0.0191)       & (0.0193)    \\ 
&    &   &   &  &  \\ 
        Observations   & 8,278 & 8,278 & 8,278 & 8,278 & 8,278\\ 
        R-squared & 0.267 & 0.236 & 0.240 & 0.217 & 0.217   \\ 
\hline 
\end{tabular}
\vspace{1mm}
\begin{minipage}{0.85\linewidth}
 {\footnotesize Note: Multigenerational estimates from eq. (\ref{eq:child-parent-gp-regression}), the dependent variable is the child's years of education (G1). Column 1 controls for the average years of education of the parents (G2) and the average years of education of all four grandparents (G3). Columns 2 and 3 control for the father's education (G2) and either the average years of education of the paternal or maternal lineage (G3). Columns 4 and 5 control for the mother's education (G2) and either the average years of education of the maternal or paternal lineage (G3). Standard errors are clustered at the family level. *** p$<$0.01, ** p$<$0.05, * p$<$0.1.}
\end{minipage}
\end{table}

\paragraph*{Family-based assortative mating. }

We so far assumed that assortative mating is ``direct'', occurring between the spouses themselves. However, in the IFLS, 28\% of women and 19\% of men born between the 1920s and the 1960s report that their parents chose their spouse. This practice varies by socioeconomic background, with the likelihood of parental involvement being negatively correlated with socio-economic status.\footnote{For each additional year of education of the wife, the probability that her partner was selected by her parents decreases by about 1.2 percentage points; for each additional log point of parental earnings, the probability decreases by about 3 percentage points.} We also observe considerable variation across provinces. In South Sulawesi, more than 75\% of women report parental involvement in spousal selection, whereas in South Sumatra, only about 6\% do so. This variation in the \textit{structure} of assortative processes reflects the differential ethnic composition across provinces (\citealt{AshrafBauNunnVoena2020}). 

What are the implications of such ``family-based'' assortative matching? To illustrate its potential effects on inter- and multigenerational mobility, we assume that the assortative process can be represented by the linear projection
\begin{equation}
e_{i,t-1}^{x} =me_{i,t-2}^{y,m}+v_{i,t-1}\quad\text{ for \ensuremath{x=\left\{ m,p\right\} }, \ensuremath{y=\left\{ m,p\right\} } and \ensuremath{y \neq x}}\label{eq:latent_family_AM}
\end{equation}
i.e. we assume that the mothers of each spouse in the parent generation (G2) ``choose'' the respective partner for their child. As shown in Appendix \ref{sec:familyAM_Appendix}, this assortative process implies the spousal correlation $Cov(e_{it-1}^{p},e_{it-1}^{m})=m\lambda_{I}$,
where $\lambda_{I}=\frac{\tilde{\lambda}}{2-m\tilde{\lambda}}$. For a given matching parameter $m$, the spousal correlation under family-based sorting tends to be smaller than the corresponding correlation in the model with ``direct'' assortative mating (as $\lambda_{I} \leq 1$), an implication that we can test in our data.

Moreover, for given values of $\{\rho,\lambda,m\}$ the intergenerational correlation $\beta_{-1}$ is smaller than the corresponding moment in the latent factor model with ``direct'' assortative mating, while the three-generation correlation $\beta_{-2}$ is larger (see Appendix \ref{sec:familyAM_Appendix}). Unless $m=\tilde{\lambda}=1$, the ratio $\frac{\beta_{-2}}{\beta_{-1}}=\frac{\tilde{\lambda}+m\left(2-m\tilde{\lambda}\right)}{2}$ is therefore necessarily larger than
the corresponding ratio in the latent factor model with ``direct''
assortative mating (which equals $\frac{\beta_{-2}}{\beta_{-1}}=\frac{\tilde{\lambda}+m\tilde{\lambda}}{2}$). Moreover, the grandparent coefficient in a regression of offspring status on parent and grandparent status from the \textit{same} lineage (i.e., father and paternal grandparent, or mother and maternal grandparent), equals
\begin{equation}
\beta_{gp}=\frac{\beta_{-2}-\beta_{-1}^{2}}{1-\beta_{-1}^{2}}=\frac{\rho^{2}\tilde{\lambda}\frac{\lambda_{I}+m}{2}-\rho^{4}\lambda_{I}^{2}}{1-\rho^{4}\lambda_{I}^{2}}\label{eq:betagp_familyAM}
\end{equation}
which is larger than $\beta_{gp}$ in the latent factor model with
direct assortative mating in eq. (\ref{eq:betagp_directAM}),
because $\beta_{-2}$ is larger and $\beta_{-1}$ is smaller.%

The family-based assortative mating considered here therefore implies
\emph{stronger} multigenerational transmission. The intuition is that
if the parents choose their child's spouse, the characteristics of that spouse will tend to depend not only on the child's own but also the parents' traits. This introduces another source of persistence that is not fully captured by conventional parent-child correlations. However, this prediction hinges on the specific structure of the assortative process. While we assumed matching on parents' endowments $e$, alternative structures are also plausible. The broader takeaway from our discussion,  therefore, is not a definitive prediction about the sign of $\beta_{gp}$, but rather the understanding that this coefficient -- and the relative strength of
inter- and multigenerational correlations -- depends on the structure
of the assortative process, and will therefore differ between countries
or groups following different marital customs.

\subsection{Evidence on family-based assortative mating and multigenerational transmission}\label{subsec:familyAM_evidence}

To empirically assess whether the structure of the marriage market in Indonesia impacts multigenerational patterns, we exploit the rich information collected in the IFLS about marital customs. As mentioned, a large fraction of individuals in the second generation of our sample declares that the spouse was selected by their family, a practice which is more common among the less educated. We first investigate whether spousal correlations in education vary depending on who selected the spouse, and then compare inter- and multigenerational coefficients for different groups.  

Table \ref{tab:ass2g} shows the coefficient estimates from different spousal regressions, where the dependent variable is the years of education of an individual and the independent variables are the years of education of the spouse (Column 1) and the parents of the spouse (Columns 2-3). Each variable is interacted with an indicator for family-based matching (\textit{FamilyChoice}), which equals 1 if the wife in the couple declares that her husband was selected by her parents.\footnote{The sample used in this analysis includes all couples observed in the IFLS data for whom information on the parents of both spouses and on who selected their spouse is available. Hence, this sample is larger than the one used for our main analysis, which also requires the observation of a third generation.} 

Column (1) of Table \ref{tab:ass2g} shows that, when the family selects a woman's spouse, the correlation between the years of education of the two partners (a proxy for assortative mating) is smaller. In contrast, the correlation between the husband and his father-in-law, or between the wife and her mother-in-law, is more positive (Columns 2-3). These results are consistent with the family-based assortative mating process described in Section \ref{sec:Model_ass_mating}, which implies that, in families where the parents select the spouse for their child, the spouses are more different from each other but more similar to their respective in-laws than in families where the spouses choose each other without their parents' involvement. 

\begin{table}[t]
   \singlespacing
  \centering
  \caption{Assortative Mating over Two Generations}  \label{tab:ass2g}%
    \begin{tabular}{lccc}
      \toprule

    {\it Dependent Variable} & Husband's & Husband's  & Wife's \\
    {\it  } &  Education & Education &  Education \\
    \hline
    
    {\it Spouse} & Wife & Wife & Husband \\
    {\it Parent } &  -- & Wife's Father & Husb.'s Mother \\
              & (1)   & (2)   & (3) \\
    \midrule
    Spousal Education   & 0.723***    & 0.662***    & 0.607***    \\
                        & (0.011)    & (0.012)    & (0.010)    \\
    Spousal Education   & -0.0465**   & -0.066***  & -0.140***   \\
    \ $\times 1(Family Choice)$& (0.018)    & (0.024)    & (0.019)    \\
&&& \\
    Parental Education  &               & 0.108***    & 0.203***    \\
                          &             & (0.011)   & (0.012)    \\
    Parental Education &                & 0.067**    & 0.068**    \\
    \ $\times 1(Family Choice)$&        & (0.030)    & (0.033)    \\
&&& \\
    $1(Family Choice)$ & -0.456***   & -0.496***   & 0.341***    \\
                          & (0.129)     & (0.130)     & (0.113)     \\
    Constant                & 2.744***    & 2.680***    & 1.822***    \\
                          & (0.098)     & (0.097)     & (0.085)    \\
&&&\\
Observations              & 11,875      & 11,007      & 10,595      \\
R-squared                 & 0.563       & 0.579       & 0.636       \\
    \bottomrule
     \end{tabular}%

    \vspace{2mm}

\begin{minipage}{0.7\linewidth}
 \footnotesize{Note: The dependent variable is the years of education of the husband (columns 1 and 2) or the wife (column 3). The independent variables are the years of education of the wife (column 1), the wife and her father (column 2), or the husband and his mother (column 3). The indicator \textit{FamilyChoice} equals 1 if the wife declares that her husband was selected by her family. Controlling for birth cohort of the wife fixed effects, standard errors are clustered at the level of the wife's {\it kabupaten} (district) of birth.*** p$<$0.01, ** p$<$0.05, * p$<$0.1}
\end{minipage}
\end{table}%

In Section \ref{sec:Model_ass_mating}, we showed that family-based matching would imply a $larger$ multigenerational coefficient. To test this implication, we estimate the multigenerational process in eq. (\ref{eq:child-parent-gp-regression}) when considering the average education of all ancestors by interacting each regressor with the same indicator \textit{FamilyChoice} as in Table \ref{tab:ass2g} that equals one if a woman (G2)'s family chose her spouse.\footnote{In Table \ref{tab:ass2g}, we use the wife's answer about whether her husband was chosen by her family as a proxy for traditional marital practice. The results also hold when using instead the husband's report that his wife was selected by his parents.}  Column (1)  of Table \ref{tab:ass3g} reports the estimates of the baseline coefficients and of the coefficients for this interaction. These results show that, conditional on grandparental education, parental education is less predictive of the child's education when the mother's family was involved in the choice of her spouse (the child's dad).  In contrast, grandparental education is more predictive of the child's education when the parents were matched by the grandparents. In fact, the coefficient on grandparental education is negative and  statistically significant when the parents did not select the spouse of their daughter (coefficient ``Grandparental Education''), but turns positive when they did (interaction with $Family Choice$).

While the preceding analysis treats marital arrangements as largely exogenous, parental involvement in spouse selection may itself be endogenous and correlated with unobserved household characteristics. To better proxy for social norms that are plausibly orthogonal to individual traits, we exploit cross-province and cross-ethnicity differences in the prevalence of family-arranged marriages. Province and ethnicity jointly account for nearly 20 percent of the variation in the share of mothers reporting that their spouse was selected by their family.\footnote{At the province level, this share is positively correlated with the value of bride price and dowry payments. Note that while the IFLS asks about bride price and dowry together without distinguishing between the two, the predominant marriage custom in Indonesia is bride price; the only exception among the main ethnic groups is the matrilocal Minangkabau, who practice dowry and represent 4.6\% of our sample \citep{AshrafBauNunnVoena2020}. Across IFLS ethnic groups, three of the seven groups we classify as having a high prevalence of parental spouse selection are also identified as groups that practice bride price in \citet{AshrafBauNunnVoena2020}.} We therefore classify provinces and ethnic groups according to whether parental involvement in spouse selection is prevalent and interact this indicator with parental and grandparental education in estimating the multigenerational equation's parameters. The indicator equals one if more than 40 percent of mothers and 20 percent of fathers in the relevant province or ethnic group report that their spouse was selected by their family.\footnote{Exploiting this definition, we identify 5 provinces (West Sumatra, East Java, South Kalimantan, South Sulawesi, Central Java) and 7 ethnicities (Bugis, Maduranese, Minang, Banjar, Makasar, Nias, Toraja) where the practice is more prevalent. Individuals in these provinces/ethnic groups account for 39\% and 16.5\% of the population, respectively.} Columns (2) and (3) present estimates using alternative definitions of the high family-matching indicator, constructed at the province-of-birth and ethnicity levels, respectively. The coefficients remain remarkably stable in both magnitude and statistical significance relative to Column (1). This robustness across definitions reinforces the interpretation that family-based spouse selection is associated with a larger multigenerational transmission coefficient.\footnote{In Table \ref{tab:ass_mat_lin} we show that the coefficient on the interaction between education of grandparents and assortative mating practices at the family level when estimated for different lineages is consistently positive and of magnitude comparable to the estimates in Table \ref{tab:ass3g}.}

	\begin{table}[t]
	\singlespacing
	\centering
			\caption{Multigenerational Transmission and Assortative Practices}\label{tab:ass3g}
			\begin{tabular}{l*{3}{c}}
				\toprule
				&\multicolumn{1}{c}{(1)} & \multicolumn{1}{c}{(2)} & \multicolumn{1}{c}{(3)} \\
				&\multicolumn{1}{c}{Individual} & \multicolumn{1}{c}{Province} & \multicolumn{1}{c}{Ethnicity} \\
				&\multicolumn{1}{c}{Definition} & \multicolumn{1}{c}{Definition} & \multicolumn{1}{c}{Definition} \\
				\midrule
				Parental Education        & 0.515$^{***}$ & 0.488$^{***}$ & 0.500$^{***}$ \\
				& (0.0161)      & (0.0182)      & (0.0153)      \\
				Parental Education      & -0.0352 & 0.0425 & 0.0273 \\
				 \ $\times 1(FamilyChoice)$ & (0.0381) & (0.0293) & (0.0394) \\
				&&&\\
				Grandparental Education & -0.0680$^{**}$ & -0.0799$^{***}$ & -0.0648$^{**}$ \\
				& (0.0265)      & (0.0303)      & (0.0254)      \\
				Grandparental Education  & 0.114$^{*}$ & 0.0956$^{**}$ & 0.136$^{**}$ \\
				 \ $\times 1(FamilyChoice)$  & (0.0609) & (0.0478) & (0.0667) \\
                 \\
				$1(FamilyChoice)$ & 0.175         & -0.383$^{**}$ & -0.275        \\
				& (0.176)       & (0.161)       & (0.221)       \\
				\\
                Constant            & 6.976$^{***}$ & 7.194$^{***}$ & 7.095$^{***}$ \\
				& (0.102)       & (0.108)       & (0.0872)      \\
				Observations        & 8,161 & 8,278 & 8,278 \\
				R-squared           & 0.269 & 0.270 & 0.269 \\
				\bottomrule
			\end{tabular}
        \vspace{2mm}

\begin{minipage}{0.86\linewidth}
 \footnotesize{Note: The table reports the estimated parameters of different variants of eq. (\ref{eq:child-parent-gp-regression}). The indicator \textit{FamilyChoice} equals 1 if the mother declares that her husband was selected by her family in Column (1), if the child is born in a province where the practice of the spouse being selected by the family is more prevalent in Column (2) and if the child belongs to an ethnicity where the practice of the spouse being selected by the family is more prevalent in Column (3). Because of some missing answers to the question of who selected the spouse, the sample used in Column (1) is slightly smaller. Grandparental education is the average of the years of education completed by all the four grandparents. Standard errors are clustered at the family level. *** p$<$0.01, ** p$<$0.05, * p$<$0.1}. 
\end{minipage}
    \end{table}

While other mechanisms may be at play, the evidence presented here aligns with the theory outlined in Section \ref{sec:Model_ass_mating}. In particular, it suggests that a common practice in Indonesian household formation -- parental selection of children's spouses -- contributes to the long-term persistence of educational outcomes across generations. Declines in this practice over time (the fraction of women declaring that their partner was selected by their parents was more than 20\% in the early 1960s and around 7\% for the most recent generations) will thus also affect the pattern of multigenerational transmission. However, while these results suggest that family-based matching affects the multigenerational transmission process, they do not explain our baseline finding in Section \ref{sec:correlations} that the grandparent-child coefficient is small or even negative in Indonesia.

\section{Conclusion} \label{sec:conclusion} 

Economic mobility across generations is a central policy concern, especially in developing countries where mobility is low and economic hardship constrains opportunities for social and economic advancement. Factors thought to limit mobility include financial and credit constraints, labor market segmentation, information frictions \citep{Piraino:2021}, strong assortative mating \citep{torche2015}, low public investment in education and health \citep{narayan2018fair}, and weaker institutions and governance \citep{Bourguignon2007}. However, while we have much evidence on \textit{inter}generational mobility from one generation to the next, much less is known about the persistence of socioeconomic advantages over multiple generations. Our central argument is that multigenerational dynamics are bound to differ between the developing and developed world -- as well as across developing countries -- so that the observation of high intergenerational persistence in developing countries need not imply strong long-run persistence.

To illustrate this argument, we used complete educational histories spanning three generations to first show that multigenerational correlations decay more rapidly in Indonesia than in high-income countries. To understand why, we compared the multigenerational implications of different transmission mechanisms in standard models of intergenerational mobility. We showed that mechanisms such as direct parental investments, credit constraints, and marital customs affect not only the parent-child correlation in economic status, but also have distinct dynamic implications. Because these mechanisms may be more salient or operate differently in developing countries, multigenerational dynamics are likely to differ as well. 

Our rich survey data allowed us to study some of these mechanisms, and their implications for multigenerational transmission, in more detail. We first considered the role of direct parental investments and financial constraints. In the Becker-Tomes model, the coefficient on grandparent status in a child-parent-grandparent regression should be more negative when financial constraints are more important. Consistent with this prediction, we found that this ``grandparent coefficient'' is more negative for boys in areas where expenditures on education are higher. Going beyond descriptive associations, we leveraged plausibly exogenous variation in financial constraints induced by the 1997 Asian Financial Crisis, showing that the coefficient is lower for boys who lived in regions that were more negatively affected by the downturn. Binding financial constraints could thus be one reason for the multigenerational patterns observed in Indonesia. 

Marital customs are another key factor influencing multigenerational dynamics. Recognizing that approximately 28\% of women and 19\% of men in our sample reported having their spouses chosen by their parents, we derived the theoretical implications of such ``family-based'' assortative mating. When a woman's family selects her spouse, we expect a weaker correlation in education between spouses, a stronger correlation between an individual and their parents-in-law, and overall higher multigenerational persistence. Exploiting variation in reported marital customs across Indonesian households, we found support for all three predictions. Family-based marital sorting, as is common in many developing countries, thus generates distinct multigenerational patterns.

Overall, these findings underscore the complexity of multigenerational dynamics, which reflect the relative strengths of different mechanisms. Our work is of course far from conclusive in identifying the factors that shape long-run persistence. And while our analysis benefited from the availability of rich survey data in Indonesia, our specific results may not directly translate to other contexts. Nonetheless, our findings highlight why multigenerational dynamics will follow distinct patterns in developing countries and point to potential mechanisms to understand these patterns.

\newpage
\clearpage
\singlespacing
\bibliographystyle{aea}
\bibliography{References}

\newpage 

\appendix
\counterwithin{table}{section}

\section{Appendix}

\singlespacing
\setcounter{table}{0}
\setcounter{figure}{0}
\renewcommand{\thetable}{A\arabic{table}}
\renewcommand{\thefigure}{A\arabic{figure}}
\pagenumbering{roman}
\setcounter{footnote}{0}

\renewcommand*{\thefootnote}{\arabic{footnote}}

\setstretch{1.5}
\subsection{The grandparent coefficient under non-stationarity}\label{sec:GP-Coeff-Nonstationary}

In a multivariate regression of child outcome $y_{it}$ on parent outcome $y_{it-1}$ and grandparent outcome $y_{it-2}$, the coefficient on the latter is positive if and only if the iteration of parent-child coefficients understates the observed persistence across three generations. 

Without assuming stationarity, the grandparent coefficient equals (Frisch-Waugh-Lovell theorem)
\begin{equation}
\beta_{gp}=\frac{Cov(y_{it},\tilde{y}_{it-2})}{Var(\tilde{y}_{it-2})}
\end{equation}
where $\tilde{y}_{it-2}$ is the residual from regressing $y_{it-2}$
on $y_{it-1}$, i.e.
\[
\tilde{y}_{it-2}=y_{it-2}-\frac{Cov(y_{it-1},y_{it-2})}{Var(y_{it-1})}y_{it-1}
\]
We therefore have
\begin{align}
\beta_{gp} &=\left(\frac{Cov(y_{it},y_{it-2})}{Var(y_{it-2})}-\frac{Cov(y_{it-1},y_{it-2})}{Var(y_{it-1})}\frac{Cov(y_{it},y_{it-1})}{Var(y_{it-2})}\right)\frac{Var(y_{it-2})}{Var(\tilde{y}_{it-2})} \nonumber \\ 
 &=\left(\beta_{-2}-\beta_{-1}^{gp\rightarrow p}\beta_{-1}^{p\rightarrow c}\right)\frac{Var(y_{it-2})}{Var(\tilde{y}_{it-2})}
\end{align}
where $\beta_{-1}^{gp\rightarrow p}$ and $\beta_{-1}^{p\rightarrow c}$ are the two-generational slope coefficient in a regression of parent on grandparent, or child on parent outcome, respectively. We have $\beta_{gp}>0$ if and only if $\beta_{-2}>\beta_{-1}^{gp\rightarrow p}\beta_{-1}^{p\rightarrow c}$. %

\subsection{The grandparent coefficient in the Becker-Tomes model}\label{sec:Proof_BT_GP}

To derive the grandparent coefficient in the Becker-Tomes model we first derive the parent-child correlation $\beta_{-1}$ and grandparent-child correlation $\beta_{-2}$, and then use the ``duality'' result in eq. (\ref{eq:Duality_result}) in the main text. We assume that the model is stationary and normalize the variance of $e$ to one. Given eqs. (\ref{eq:LF-e}) and (\ref{eq:BT-y}),  
the intergenerational correlation equals
\begin{align}
\beta_{-1} & =\frac{Cov(y_{it,}y_{it-1})}{Var(y_{it-1})} \nonumber  \\
 & =\frac{Cov(\gamma y_{it-1}+\rho e_{it},y_{it-1})}{Var(y_{it-1})} \nonumber  \\
 & =\gamma+\rho\lambda \frac{Cov(e_{it-1},y_{it-1})}{Var(y_{it-1})} \nonumber  \\
 & =\gamma+\frac{\rho^{2}\lambda}{(1-\gamma\lambda)Var(y_{it-1})}
\end{align}
where $Cov(e_{it-1},y_{it-1})=Cov(e_{it-1},\gamma y_{it-2}+\rho e_{it-1})=\lambda\gamma Cov(e_{it-2},y_{it-2})+\rho=\rho/(1-\gamma\lambda)$, as in steady state $Cov(e_{it-1},y_{it-1})=Cov(e_{it-2},y_{it-2})$. 

The grandparent-child correlation equals
\begin{align}
\beta_{-2} & =\frac{Cov(y_{it,}y_{it-2})}{Var(y_{it-2})} \nonumber  \\
 & =\frac{Cov(\gamma y_{it-1}+\rho e_{it},y_{it-2})}{Var(y_{it-2})} \nonumber  \\
 & =\gamma \frac{Cov(y_{it-1,}y_{it-2})}{Var(y_{it-2})}+\rho \frac{Cov(e_{it},y_{it-2})}{Var(y_{it-2})} \nonumber  \\
 & =\gamma\beta_{-1}+\frac{\rho^{2}\lambda^{2}}{(1-\gamma\lambda)Var(y_{it-2})}
\end{align}

Exploiting the ``duality'' result in eq. (\ref{eq:Duality_result}) it follows that
\begin{align}
\beta_{gp} =\frac{\beta_{-2}-\beta_{-1}^{2}}{1-\beta_{-1}^{2}}\nonumber 
\end{align}

Normalizing the variance of $y$ to one, we thus have
\begin{align}
 \beta_{gp} & =\frac{\gamma\left(\gamma+\frac{\rho^{2}\lambda}{1-\gamma\lambda}\right)+\frac{\rho^{2}\lambda^{2}}{1-\gamma\lambda}-\left(\gamma+\frac{\rho^{2}\lambda}{1-\gamma\lambda}\right)^{2}}{1-\left(\gamma+\frac{\rho^{2}\lambda}{1-\gamma\lambda}\right)^{2}}\nonumber \\
 & =\frac{\gamma\frac{\rho^{2}\lambda}{1-\gamma\lambda}+\frac{\rho^{2}\lambda^{2}}{1-\gamma\lambda}-2\gamma\frac{\rho^{2}\lambda}{1-\gamma\lambda}-\left(\frac{\rho^{2}\lambda}{1-\gamma\lambda}\right)^{2}}{1-\left(\gamma+\frac{\rho^{2}\lambda}{1-\gamma\lambda}\right)^{2}}\nonumber \\
 & =\frac{\frac{\rho^{2}\lambda}{1-\gamma\lambda}\left(\lambda-\gamma-\frac{\rho^{2}\lambda}{1-\gamma\lambda}\right)}{1-\left(\gamma+\frac{\rho^{2}\lambda}{1-\gamma\lambda}\right)^{2}}\label{eq:Proof_BT_GP}
\end{align}
which is eq. (\ref{eq:BT-original-betagp}) in the main text. The denominator is non-negative, as $\beta_{-1}=\gamma+\frac{\rho^{2}\lambda}{1-\gamma\lambda}\leq1$ by the definition of correlations. The coefficient can be written more compactly as
\[
\beta_{gp}=\frac{\left(\beta_{-1}-\gamma\right)\left(\lambda-\beta_{-1}\right)}{1-\beta_{-1}^{2}}.
\]
Because $\beta_{-1}-\gamma=\frac{\rho^{2}\lambda}{1-\gamma\lambda}>0$ (for $\rho, \lambda > 0$), the grandparent coefficient has the same sign as $\lambda-\beta_{-1}$. This expression nests the special cases from Section \ref{sec:Model_main}: for $\gamma=0$ it reduces to the latent factor result, while for $Var(u)=0$ we have
$\beta_{-1}=(\gamma+\lambda)/(1+\gamma\lambda)$ and $\beta_{gp}=-\gamma\lambda$.

To understand how $\beta_{gp}$ varies with $\gamma$, we cannot normalize the variance of $y$, since this variance itself depends on $\gamma$. More generally, its steady-state variance can be derived as $Var(y_{it})=(\rho^2+\sigma^2+2\gamma\rho^2\lambda/(1-\gamma\lambda))/(1-\gamma^2)$, where $\sigma^2$ denotes the variance of the error term $u_{it}$ in eq. (\ref{eq:BT-y}). The parameter $\sigma^2$ did not appear in the expressions in the main text, as it is implicitly determined by the other parameters when $y$ is normalized. In this case, the grandparent coefficient is given by (the following steps have been derived using the software \textit{Mathematica})
\begin{align}
\beta_{gp} &= \frac{\lambda \rho^2 \left(\gamma^2 \lambda \sigma^2+\gamma \left(\lambda^2-1\right) \rho^2-\gamma \left(\lambda^2+1\right) \sigma^2+\lambda \sigma^2\right)}{-2 \rho^2 \sigma^2 (\gamma \lambda-1)+\sigma^4 (\gamma \lambda-1)^2-\left(\lambda^2-1\right) \rho^4}.\label{eq:beta_gp_BT_general}
\end{align}
Its derivative with respect to $\gamma$ is
\begin{align}
\frac{\partial\beta_{gp}}{\partial\gamma} &= \frac{- \lambda \rho^2 \left(\rho^2 \sigma^4 \left(\gamma^2 \left(\lambda^4+\lambda^2\right)-4 \gamma \lambda-\lambda^2+3\right)+\left(\lambda^2-1\right) \rho^4 \sigma^2 \left(2 \gamma \lambda-\lambda^2-3\right)\right)}{\left(2 \rho^2 \sigma^2 (\gamma \lambda-1)-\sigma^4 (\gamma \lambda-1)^2+\left(\lambda^2-1\right) \rho^4\right)^2} \nonumber \\
& \ \ \ \  + \frac{\lambda \rho^2 \left(\left(\lambda^2-1\right) \sigma^6 (\gamma \lambda-1)^2-\left(\lambda^2-1\right)^2 \rho^6\right)}{\left(2 \rho^2 \sigma^2 (\gamma \lambda-1)-\sigma^4 (\gamma \lambda-1)^2+\left(\lambda^2-1\right) \rho^4\right)^2},
\end{align}
which is negative (if $\lambda > 0$, $\rho > 0$). In contrast, the sign of the derivative of $\beta_{gp}$ with respect to $\sigma$ depends on parameter values and can be either positive or negative.

\subsection{The grandparent coefficient with direct assortative mating}
\label{sec:latent_lineages}

In this Appendix, we show how the grandparent coefficient varies in a latent factor model with assortative mating, depending on which lineages are included in the analysis. Assume that assortative mating is ``direct'', as specified in eq. (\ref{eq:latent_direct_AM}). The grandparent coefficient in a regression of offspring status on parent and grandparent status from the \textit{same} lineage (i.e., father and paternal grandparent on the paternal lineage $p$, or mother and maternal grandparent on the maternal lineage $m$)
\begin{equation}
y_{it}=\beta_{p}y_{it-1}^{x}+\beta_{gp}y_{it-2}^{x,y}+\epsilon_{it}\quad\text{ for \ensuremath{x=\left\{ m,p\right\} }, \ensuremath{y=\left\{ m,p\right\} }}\label{eq:GP_coefficient_same_lineage_APP}
\end{equation}
then still equals
\begin{equation}
\beta_{gp}=\frac{\beta_{-2}-\beta_{-1}^{2}}{1-\beta_{-1}^{2}}=\frac{\rho^{2}\lambda^{2}-\rho^{4}\lambda^{2}}{1-\rho^{4}\lambda^{2}},\label{eq:betagp_directAM_APP}
\end{equation}
in line with our results for the one-parent version of the latent factor model, but with the parameter $\lambda$ now also reflecting the strength of the assortative process (see main text).

Conversely, the grandparent coefficient in a regression of offspring status on parent and grandparent status from \textit{different} lineages (i.e., father and maternal grandparent, or mother and paternal grandparent),
\begin{equation}
y_{it}=\beta_{p}^{\prime}y_{it-1}^{x}+\beta_{gp}^{\prime}y_{it-2}^{y,z}+\epsilon_{it}\quad\text{ for \ensuremath{y\neq x} and \ensuremath{z=\left\{ m,p\right\} }}\label{eq:GP_coefficient_diff_lineage_APP}
\end{equation}
equals
\begin{equation}
\beta_{gp}^{\prime}=\frac{\beta_{-2}-\beta_{-1}\beta_{-1}^{\prime}}{1-\left(\beta_{-1}^{\prime}\right)^{2}}=\frac{\rho^{2}\lambda^{2}-m\rho^{4}\lambda^{2}}{1-m^{2}\rho^{4}\lambda^{2}},
\end{equation}
implying $\beta_{gp}^{\prime}>\beta_{gp}$ if assortative mating is
imperfect ($0\leq m<1$). We therefore expect the grandparent coefficient
to be less positive when including grandparent(s) from the \emph{own}
lineage (e.g., including father and paternal grandmother/grandparents),
and more positive when including a parent from the \emph{other} lineage
(e.g. father and maternal grandmother/grandparents). 

Finally, the grandparent coefficient in a regression on the status
of grandparent and \textit{both} parents, 
\begin{equation}
y_{it}=\beta_{x}y_{it-1}^{x}+\beta_{y}y_{it-1}^{y}+\beta_{gp}^{\prime\prime}y_{it-2}^{x,z}+\epsilon_{it}\quad\text{ for \ensuremath{y\neq x} and \ensuremath{z=\left\{ m,p\right\} }}\label{eq:GP_coefficient_both_parents_APP}
\end{equation}
equals $\beta_{gp}^{\prime\prime}=Cov(y_{it},\tilde{y}_{it-2}^{x,z})/Var(\tilde{y}_{it-2}^{x,z}),$
where $\tilde{y}_{it-2}^{x,z}$ is the residual from regressing $y_{it-2}^{x,z}$
on $y_{it-1}^{x}$ and $y_{it-1}^{y}$. The slope coefficients in this
auxiliary regression equal $\left(\rho^{2}\lambda-m^{2}\rho^{4}\lambda\right)/\left(1-m^{2}\rho^{4}\right)$
on $y_{it-1}^{x}$ and $\left(m\rho^{2}\lambda-m\rho^{4}\lambda\right)/\left(1-m^{2}\rho^{4}\right)$
on $y_{it-1}^{y}$. After simplification, we have 
\begin{equation}
\beta_{gp}^{\prime\prime}=\frac{\rho^{2}\lambda^{2}(\rho^{2}-1)(m\rho^{2}-1)}{1-m^{2}\rho^{4}+\rho^{4}\lambda^{2}\left(m^{2}(2\rho^{2}-1)-1\right)}
\end{equation}
where $\beta_{gp}^{\prime\prime}<\beta_{gp}^{\prime}$  if $0<\rho<1$,
$0\leq m\leq1$, and $0<\lambda\leq1$. The numerator is necessarily
positive, because both $(\rho^{2}-1)$ and $(m\rho^{2}-1)$ are
negative (as $\rho<1$ and $m<1$). The denominator is also positive,
so $\beta_{gp}^{\prime\prime}$ is necessarily positive in this model, similarly as $\beta_{gp}$ is positive in the one-parent latent factor
model.

\subsection{Family-based assortative mating}
\label{sec:familyAM_Appendix}

We next derive the grandparent coefficient under ``family-based'' assortative matching. Assume that the assortative process can be represented by eq. (\ref{eq:latent_family_AM}), that is, we assume that the mothers of each spouse in the parent generation (G2) ``choose'' the respective partner for their child. This assortative process implies the spousal
correlation $Cov(e_{it-1}^{p},e_{it-1}^{m})=m\lambda_{I}$,
where $\lambda_{I}=\frac{\tilde{\lambda}}{2-m\tilde{\lambda}}$.\footnote{Note that $Cov(e_{it-1}^{p},e_{it-1}^{m})=Cov(me_{i,t-2}^{m,m},\tilde{\lambda}(e_{i,t-2}^{m,m}+e_{i,t-2}^{m,p})/2)=m\tilde{\lambda}\left(1+Cov(e_{i,t-2}^{m,m},e_{i,t-2}^{m,p})\right)/2=m\tilde{\lambda}/(2-m\tilde{\lambda}),$
where the last step follows because in steady state we have $Cov(e_{it-1}^{p},e_{it-1}^{m})=Cov(e_{i,t-2}^{m,m},e_{i,t-2}^{m,p})$.} 

The intergenerational correlation is now given as\footnote{The corresponding correlations in $e$ are $Cov(e_{i,t},e_{i,t-1}^{m})=Cov(\tilde{\lambda}(e_{i,t-1}^{m}+e_{i,t-1}^{p})/2,e_{i,t-1}^{m})=\tilde{\lambda}\left(1+Cov(e_{i,t-1}^{m},e_{i,t-1}^{p})\right)/2=\frac{\tilde{\lambda}}{2-m\tilde{\lambda}}=\lambda_{I}$,
which for $m>0$ again tends to be smaller than the corresponding
moment in the latent factor model with ``direct'' assortative mating
(in which $Cov(e_{i,t},e_{i,t-1}^{m})=\tilde{\lambda}\left(1+m\right)/2$).
Following similar steps, we can derive $Cov(e_{it},e_{it-2}^{m,m})=\frac{\tilde{\lambda}^{2}+m\tilde{\lambda}\left(2-m\tilde{\lambda}\right)}{4-2m\tilde{\lambda}}$.}
\[
\beta_{-1}=\rho^{2}\lambda_{I},
\]
which, for given values of $\{\rho,\lambda,m\}$, is \emph{smaller}
than the corresponding moment in the latent factor model with ``direct''
assortative mating (as $\lambda_{I}\leq\lambda$, see above), while the three-generation correlation is
\[
\beta_{-2}=\rho^{2}\frac{\tilde{\lambda}^{2}+m\tilde{\lambda}\left(2-m\tilde{\lambda}\right)}{4-2m\tilde{\lambda}},
\]
which, for given values of $\{\rho,\lambda,m\}$, is \emph{larger} than
the corresponding moment in the latent factor model with ``direct''
assortative mating. Unless $m=\tilde{\lambda}=1$, the ratio
\[
\frac{\beta_{-2}}{\beta_{-1}}=\frac{\tilde{\lambda}+m\left(2-m\tilde{\lambda}\right)}{2}
\]
is therefore necessarily larger than
the corresponding ratio in the latent factor model with ``direct''
assortative mating (which equals $\frac{\beta_{-2}}{\beta_{-1}}=\frac{\tilde{\lambda}+m\tilde{\lambda}}{2}$).

What are the implications for the grandparent coefficient in a multigenerational regression? The grandparent coefficient in a regression of offspring status on parent and grandparent status from the \textit{same} lineage (i.e., father and paternal grandparent, or mother and maternal grandparent), as in eq. (\ref{eq:GP_coefficient_same_lineage_APP}),
equals
\begin{equation}
\beta_{gp}=\frac{\beta_{-2}-\beta_{-1}^{2}}{1-\beta_{-1}^{2}}=\frac{\rho^{2}\tilde{\lambda}\frac{\lambda_{I}+m}{2}-\rho^{4}\lambda_{I}^{2}}{1-\rho^{4}\lambda_{I}^{2}}\label{eq:betagp_familyAM}
\end{equation}
which is larger than $\beta_{gp}$ in the latent factor model with
direct assortative mating in eq. (\ref{eq:betagp_directAM_APP}),
because $\beta_{-2}$ is larger and $\beta_{-1}$ is smaller.%

\subsection{Detailed information on key variables from the IFLS}\label{sec:varinfo}

To construct a dataset containing information on three generations, we exploit the longitudinal nature of the IFLS. Within households, all individuals who appear in one survey are followed up in subsequent surveys. If an individual exits the original household and forms a new household, this individual and all members of the new household are interviewed in the next wave. This is true even if individuals move to a province which is not included in the original survey design. When individuals move to a new household, they retain their individual-level identification number, allowing us to track, for example, individuals who were children in earlier waves and form new households in later waves. Given the long time span of the surveys (21 years), we have information on individuals from their childhood into their adulthood.

\paragraph{Education.} 

 The panel nature of the dataset implies that we potentially have information on the education of an individual at a maximum of five different points in time and from different questions at each point in time. {\color{blue} Within each wave, some individuals appear in the education files with multiple records corresponding to different schooling levels they attended. In these cases, we assign the highest completed education level recorded in that wave.} Across waves, we take advantage of the panel structure of the data to check for consistency in the education variable. In some cases, we observe inconsistency in the reported education across survey waves. For example, it is possible that a person reports having a high school degree in 1993 and then reports having only completed elementary school in 1997, and it is also possible that a person reports having a high school degree in one wave but the head of the household states that the same person only has elementary education. In these cases of inconsistency across waves{\color{blue}, we take the mode of education reported in all the waves where the person is above 25.} 
  
  To construct the measure for the parents and grandchildren,  we use the following procedure.  If an individual first enters the IFLS at 25 years of age or more, we keep the level of education reported in that wave as his highest level of education, as long as the information is coherent within that wave.\footnote{We have also considered checking consistency not only within the first wave but also across all the waves a person appears in. We prefer using the information in the first observation because this is the closest to age 25. Using the mode across waves instead of the first observation, when there is inconsistency across waves gives slightly higher average education for all.} %
If an individual appears in multiple waves, with the first appearance at or below 24 years of age, we take the education reported in the first wave where the person is above 25 as the highest level of education attained, as long as the reported education level is increasing across waves prior to reaching 25 and is consistent with the first wave where the person is above 25.

To construct the education measure of grandparents, we use two sources of information. If the grandparent is still alive and resides in the same household as the parent, then we have both the self-reported education as well as the level reported by the heads of household. Otherwise, we only have information on the education of grandparents as reported by the heads of household. In practice, co-resident grandparents are a minority. To have a consistent measure, we prefer to use education as reported by the head of the household even for co-resident grandparents.

\paragraph{Individual demographics.} %
 We assign to each individual the year of birth reported in the most recent wave where he/she is present, assuming this is the most accurate information in case there is any inconsistency surrounding the reported year of birth. Some household members are also asked about the year of birth of their parents: we exploit this information to generate a variable reporting the year of birth of the grandparents in our three-generation dataset. We check for data consistency by examining the distribution of the years of birth of the grandparents and parents (see Section~\ref{sec:descriptives}). 

\paragraph{Province/district of birth and migration.} Each wave of the survey elicits information on place of birth, residence, and relocation. Province of birth is consistent across waves for the vast majority of individuals in our sample. When not reported, we include as place of birth the place where the individual is observed residing in the first wave of the survey. Information on place of residence is reported from age 12.\footnote{With this information, we have checked that only a small number of individuals were subject to the ``transmigration'' policy (more details on the policy are discussed in \cite{bazzi2016}).}

\paragraph{Marital traditions.} The surveys elicit information on the mechanism of marriage formation. In particular, we focus our attention on the question about who in the household selects the spouse. Possible answers include ``parents'', ``self'' or ``family''.\footnote{Note that more questions about marital traditions are asked in the survey. In particular, we have information about the dowry type and amount and the number of wives for individuals who have more than one. However, since not all individuals answer the question about dowry it is not possible to discriminate whether individuals did not pay for dowry or whether it is just missing information. The cases of polygamy are very few.}

\paragraph{Education expenditures.} Each household is asked the total school-related expenditures during the relevant school year. Specifically, households are asked the amount they spent on school fees (including registration, examinations), school supplies (including books, uniforms), and logistics (including transportation, food and housing costs). We want to capture the per-child expenditures in education only for the years when third generation children are in school. Because their average years of schooling is less than 10, for each cohort we want to use information in the waves when they are between age 15 and age 6 (the school starting age). We thus consider information in the 1993 wave (the first wave we observe) for all children born before 1982, in the 1997 wave for those born between 1982 and 1984 and in the 2000 wave for those born between 1985 and 1988. However, the 1993 wave asks different questions compared to the following waves.\footnote{In particular, in the 1993 survey the following questions are included: ``Approximately what was the total monthly expenditures (e.g., tuitions, pocket money)?"; and ``Approximately what was the total of annual expenditures beyond monthly expenditures (like for school uniforms, school supplies, registration fees)?". The equivalent questions in the subsequent waves are formulated differently: ``Approximately what was the total expenditures (e.g., tuition, registration fees, exam costs) during the past year?"; and ``Approximately what was the total of annual expenditures (like for school uniforms, school supplies) during the past year?".} The resulting distribution of school expenditures is very different in the different waves. %
In the analysis in Section \ref{sec:credit}, where we compute the expenditure share at the province level, we consider the average expenditure within each province only including the expenditures for children born in 1982 and beyond. To calculate the share, we divide this number by the average earnings of parents of the same children in the same years. We adjust expenses for inflation using the consumer price index at the national level and express all measures in 2015 terms.

\paragraph{Earnings.} In each survey wave, all household members report their earnings or profits.\footnote{These two measures never overlap (individuals either have earnings or profits), depending on the type of job of the individual.} We can therefore create one unique measure of earnings, which include earnings for some individuals and profits for others. For some individuals we have observation of earnings in all five waves, while for others we have only one observation. As a measure, we take the average across all the available observations. We adjust earnings for inflation using the consumer price index at the national level and express all measures in 2015 terms.

\subsection{Additional materials for descriptive statistics}
\setstretch{1.5}
The IFLS has a relatively low rate of attrition - in 1997, the IFLS reinterviewed 94\% of households in the first wave of the IFLS. In cases where survey respondents moved in the intervening years, interviews were conducted at the new locations. %
The third and fourth waves have 95.3\% and 93.6\% re-interview rates with initial households respectively. 

Individuals in our sample from the third generation are equally split between men and women and across cohorts (see Figure~\ref{fig:distr_yob}). Some individuals in the third generation are siblings. %
In Table~\ref{tab:nchild}, we report the proportion of second generation individuals with different number of children and the average years of education conditional on the number of children: over 60\% of them have more than one child and the relationship between the number of children and the years of education is negative. For the second generation, the median year of birth among fathers is 1952, while for mothers it is 1957. Figure~\ref{fig:distr_yob_p} plots the estimated density of the birth years of parents, which shows that the youngest parents are born in the late 1960s. Figure \ref{fig:distr_yob_gp} plots the same estimated distribution for grandparents: they are born between the late 19th century and the late 1940s and, as expected, paternal grandfathers are the oldest group on average, while maternal grandmothers are the youngest. This is also reflected in the median year of birth, where the difference between year of birth of grandfathers and grandmothers is 9 years for the paternal grandparents and 7 years for the maternal grandparents.

\subsection{Validation of education measure} \label{sec:val}
\setstretch{1.5}

To validate {our} data cleaning procedure and to check whether the self-reported educational attainment from the IFLS are accurate, we compare our measure of education among the IFLS respondents with the level of education reported by the same cohorts in the 1995 and 2005 Indonesian population census. Figure~\ref{fig:validation} shows the average years of education for the parents (G2): comparing the measure {from the IFLS with that from} the Census. We restrict the comparison sample from the Census to include only the provinces {covered by the IFLS}.\footnote{We keep individuals born until 1980 since they are 25 in the 2005 census.} Although not perfectly overlapping, the averages are rather close for all the cohorts we are considering, suggesting that the measure of education we built is quite precise. Similarly, Figure~\ref{fig:validation_gp} show the difference between the IFLS and the 1995 or 2005 census in the average level of education and in the proportion of individuals with no education in each cohort for the grandparents (i.e. the levels of education of the parents reported by the head of the households). Both graphs confirm that the measure of education reported in the IFLS is in line with information from the Census.\

\subsection{Multigenerational regression by education expenditures at the province level}\label{app:expenditure}

As mentioned in Section \ref{sec:varinfo} one of the questions asked in the IFLS is the yearly expenditure in education for children. Using this information, we build a measure for the relevance of parental investments in the human capital accumulation process at the province level, capturing that in different areas education may be more or less expensive. We then estimate the coefficients of the multigenerational regression interacted with these measures, to investigate whether, in line with the theory, the grandparent coefficient is more negative in more expensive areas.

In particular, we define the expenditure share ($ExpSh$) as the ratio of average educational expenditure reported by families whose child (G1) was born in a given province to the average earnings of those same families, and is scaled to have mean zero. To ensure consistency in the educational expenditure question across waves, province-level expenditure shares are computed using only families with children aged 12--16 in the 1997 and 2000 waves (birth cohorts 1982--1988 – see section \ref{sec:varinfo}). Average earnings are trimmed at the 1st and 99th percentiles to mitigate the influence of outliers. The mean expenditure share across the 13 provinces we consider is 0.28 with a s.d. of 0.13. 

We then interact this number, as well as an indicator for this expenditure  being above 40\% in the province of interest, with both the parental and the grandparental education in the multigenerational regression eq. (\ref{eq:child-parent-gp-regression}). Two provinces, DI Yogyakarta and DKI Jakarta have values of this expenditure share above 40\%, the values of all the others provinces are smoothly distributed between 12\% and 33\%. We estimate the parameters of this modified  multigenerational equation:
\begin{equation}
y_{it,k}= \beta_{p} y_{it-1} + \beta_{p,exp}ExpSh_{k}\times y_{it-1}+
\beta_{gp} y_{it-2} + \beta_{gp,exp}ExpSh_{k}\times y_{it-2}+ \beta_{exp}ExpSh_{k}+\varepsilon_{it,k}.\label{eq:multig_expenditure}
\end{equation}
where {\it ExpSh} is the expenditure share measure, $k$ is the province, and $y_{it-1}$ and $y_{it-2}$ are the average education of parents and grandparents, respectively. Estimates of parameters $\beta_p$, $\beta_{p,exp}$, $\beta_{gp}$ and $\beta_{gp,exp}$, overall and separately by gender, are reported in Table \ref{tab:area_share}. 

\clearpage
\subsection{Tables and Figures}
 
\begin{table}[h]
\centering
\caption{Education and Proportions of Generation 2 by Number of Children}\label{tab:nchild}
\begin{tabular}{lccc}
\toprule

&\multicolumn{1}{c}{Proportion}&\multicolumn{2}{c}{Years of  Education}\\
&				        & \multicolumn{1}{c}{Fathers} &\multicolumn{1}{c}{Mothers}\\
\hline
1 Child   & 35.8 & 6.05 & 5.13 \\
2 Children& 30.7 & 6.08 & 4.91 \\
3 Children& 19.3 & 6.30 & 5.01 \\
4 Children& 9.6  & 6.00 & 4.53 \\
5 Children& 3.6  & 5.85 & 3.61 \\
6 Children& 0.9  & 4.94 & 3.61 \\
7 Children& 0.2 & 6.83 & 5.67 \\
8 Children& 0.1  & 4.00 & 5.00 \\

\hline
\end{tabular}
\end{table}

  \begin{table}[H]
    \centering
            \caption{Robustness of Parent Child Correlation Across Samples \label{tab:select}} 
            {
\def\sym#1{\ifmmode^{#1}\else\(^{#1}\)\fi}
\begin{tabular}{l*{4}{c}}
\toprule
    &\multicolumn{4}{c}{Dependent Variable: Child's Education}    \\
 \hline
Missing   &\multicolumn{1}{c}{0-1 parent} &\multicolumn{3}{c}{0 parent}    \\
\cmidrule(lr){2-2}\cmidrule(lr){3-5} 
 Missing &\multicolumn{1}{c}{0-4 gp} &\multicolumn{1}{c}{0 gp}  &\multicolumn{1}{c}{1-3 gp} &\multicolumn{1}{c}{4 gp}       \\
             & (1)   & (2)   & (3)   & (4)        \\
    \midrule
Parental education&       0.467\sym{***}&       0.485\sym{***}&       0.401\sym{***}&       0.418\sym{***}\\
                    &   (0.00973)         &    (0.0119)         &    (0.0375)         &    (0.0319) \\     

&&&&\\
Mean of y           &        9.68         &         9.7         &        9.69         &        10.3               \\
Observations        &       11,858         &        8,278      &    1,148         &        1,122         \\
R-squared           &       0.251         &       0.266         &       0.165         &       0.223         \\
\hline
\end{tabular}
}

\vspace{1mm}
\footnotesize
 \begin{minipage}{0.65\linewidth}

\textit{Notes:} The dependent variable is years of education of the child. Column (1) presents results for the sample in which we allow for missing education information for up to one parent and up to four grandparents. Columns (2)-(4) restrict the sample to observations where education information is available for both parents. Column (2) further restricts to observations with complete education information for all four grandparents (the fully restricted sample). Column (3) allows for missing education information for up to three grandparents, and Column (4) allows for missing information for all four grandparents. Standard errors clustered at the family level are reported in parentheses. *** p$<$0.01, ** p$<$0.05, * p$<$0.1.
 \end{minipage}

\end{table}

 \begin{table}[H]
    \centering
            \caption{Examining Selection into the Restricted Sample \label{tab:move}} 
{
\def\sym#1{\ifmmode^{#1}\else\(^{#1}\)\fi}
\begin{tabular}{l*{5}{c}}
\toprule
    &\multicolumn{5}{c}{Restricted Sample Dummy (= 1 if full three generation)}    \\
 \hline
             & (1)   & (2)   & (3)   & (4)   & (5)   \\
    \midrule
Move dummy           &     0.00723         &      0.0113         &     0.00518         &     0.00446         &     0.00585         \\
                    &   (0.00968)         &    (0.0113)         &    (0.0112)         &    (0.0112)         &    (0.0113)         \\

Child education &                     &                     &     0.00572\sym{***}&     0.00573\sym{***}&     0.00568\sym{***}\\
                    &                     &                     &   (0.00157)         &   (0.00157)         &   (0.00157)         \\

Female                  &                     &                     &                     &     -0.0131         &     -0.0138\sym{*}  \\
                    &                     &                     &                     &   (0.00810)         &   (0.00807)         \\
 
&&&&&\\
Parental education missing & 0  & 0 to 1   & 0 to 1 & 0 to 1 & 0 to 1 \\
Grandparental education missing & $\leq3$  & $\leq3$    & $\leq3$  & $\leq3$  & $\leq3$  \\
Year of birth FE & No & No   & No   & No   & Yes\\

&&&&&\\

Mean of y           &        .878         &        .797         &        .797         &        .797         &        .797         \\
Observations        &        9,424         &       10,382         &       10,382         &       10,382         &       10,382         \\
R-squared           &       0.000         &       0.000         &       0.003         &       0.003         &       0.008         \\
\hline
\end{tabular}
}

\vspace{1mm}
\footnotesize
 \begin{minipage}{0.95\linewidth}

\textit{Notes:} The dependent variable equals one if the observation has complete education information for both parents and all four grandparents (the restricted sample), and zero otherwise. This table analyses patterns of sample selection. Column (1) conditions on observations with no missing parental education but allows up to three grandparents to have missing data. Columns (2)-(5) relax this restriction to allow up to one parent to have missing education information, while still allowing up to three grandparents to have missing data. Columns (2)-(5) sequentially add controls for the child's generation and year of birth fixed effects. The move dummy indicates whether the individual has moved at least once before the age of 40. Standard errors are clustered at the family level and reported in parentheses. *** p$<$0.01, ** p$<$0.05, * p$<$0.1.
 \end{minipage}

\end{table}

\clearpage

\begin{table}[H]
\singlespacing
\caption{Non-Linearities in Multigenerational Regression}\label{tab:nonlinearities}
\begin{tabular}{l*{5}{c}}
	\hline
 	&\multicolumn{1}{c}{Baseline} &\multicolumn{1}{c}{Quadratic} &\multicolumn{1}{c}{Quadratic} &\multicolumn{1}{c}{Quadratic} &\multicolumn{1}{c}{Dummies}\\
    	&\multicolumn{1}{c}{(1)}&\multicolumn{1}{c}{(2)}&\multicolumn{1}{c}{(3)}&\multicolumn{1}{c}{(4)}&\multicolumn{1}{c}{(5)}\\

	\hline
	Parental Education           & 0.485\sym{***} & 0.703\sym{***} & 0.486\sym{***} & 0.686\sym{***} &                     \\
	& (0.0151)       & (0.0476)       & (0.0151)       & (0.0479)       &                     \\
	Grandparental Education          & -0.0566\sym{*} & -0.0255        & 0.153\sym{*}   & 0.110          &                     \\
	& (0.0292)       & (0.0287)       & (0.0868)       & (0.0860)       &                     \\
	Parental Education$^2$       &               & -0.0157\sym{***}&                & -0.0145\sym{***}&                    \\
	&               & (0.00315)       &                & (0.00316)       &                     \\
	Grandparental Education$^2$      &               &                & -0.0238\sym{**}& -0.0156         &                     \\
	&               &                & (0.00990)      & (0.00980)       &                     \\
	1[Edu GP =Edu P]     &               &                &                &               			& -1.466\sym{***}            \\
						&               &                &                &                			& (0.338)             \\
	1[Edu GP $>$ P]		&               &                &                &                			& -2.452\sym{***}      \\
						&               &                &                &                			& (0.227)            \\
	\\
	1[Grandparents no education]    	& -0.127        & 0.0201         & 0.189          & 0.216          		& -1.787\sym{***}      \\
							& (0.128)       & (0.175)        & (0.173)        & (0.129)        		& (0.112)             \\
	1[Parents no education]	& -0.473\sym{**}& 0.0387         & -0.463\sym{**}& 0.00447        		& -0.923\sym{**}     \\
							& (0.206)       & (0.232)        & (0.206)        & (0.233)        		& (0.368)             \\
	\\
	Observations          & 8,278         & 8,278          & 8,278          & 8,278          		& 8,278               \\
	R-squared             & 0.269         & 0.272          & 0.270          & 0.273          		& 0.125\\
	\hline
\end{tabular}

    \vspace{2mm}
\begin{minipage}{.95\linewidth}
 {\footnotesize  \textbf{Note}: The table above reports the estimates of multigenerational correlations in education from eq. (\ref{eq:child-parent-gp-regression}) including quadratic terms in parental education (column 2), grandparental education (column 3) and both (column 4). Column 5 includes a dummy for parents having the same education as grandparents and a dummy for grandparents having more education than parents. All regressions control for a dummy for all grandparents having no education and a dummy for both parents having no education. Standard errors are clustered at the family level (all children with the same mother are clustered together). *** p$<$0.01, ** p$<$0.05, * p$<$0.1.}
 \end{minipage}
\end{table}

\clearpage
\singlespacing

\begin{table}[h!]
\singlespacing
\centering

        \caption{Multigenerational Transmission in Urban vs Rural Areas}   \label{tab:rural}
\begin{tabular}{lcccccc}
\toprule
                     &\multicolumn{6}{c}{Dependent Variable: Child's Education}\\
\cmidrule(lr){2-7}  
                    &\multicolumn{3}{c}{Birthplace}&\multicolumn{3}{c}{Birthplace (adjusted)}\\
 \cmidrule(lr){2-4}\cmidrule(lr){5-7}     
                    &\multicolumn{1}{c}{Urban}&\multicolumn{1}{c}{Rural}&\multicolumn{1}{c}{Interaction}&\multicolumn{1}{c}{Urban}&\multicolumn{1}{c}{Rural}&\multicolumn{1}{c}{Interaction}\\
 
                    &\multicolumn{1}{c}{(1)}&\multicolumn{1}{c}{(2)}&\multicolumn{1}{c}{(3)}&\multicolumn{1}{c}{(4)}&\multicolumn{1}{c}{(5)}&\multicolumn{1}{c}{(6)}\\
 \hline
Grandparental Education&     -0.0406         &     0.00524         &     -0.0406         &     -0.0680\sym{**} &     -0.0145         &     -0.0680\sym{**} \\
                    &    (0.0277)         &    (0.0295)         &    (0.0277)         &    (0.0334)         &    (0.0309)         &    (0.0334)         \\
                                             
 $\times$ Rural     &                     &                     &      0.0459         &                     &                     &       0.0535              \\
                                                     &                     &                     &    (0.0388)         &                     &                     &       (0.0440)                \\

Parental Education&       0.458\sym{***}&       0.530\sym{***}&       0.458\sym{***}&       0.425\sym{***}&       0.530\sym{***}&       0.425\sym{***}\\
                    &    (0.0213)         &    (0.0175)         &    (0.0213)         &    (0.0222)         &    (0.0182)         &    (0.0222)         \\

 $\times$ Rural             &                     &                     &      0.0713\sym{***} &                     &                     &    0.104\sym{***}                  \\
                                    &                     &                     &    (0.0265)         &                     &                     &         (0.0275)                \\
 
 Rural                             &                     &                     &      -1.127\sym{***}&                     &                     &    -1.172\sym{***}                 \\
                                       &                     &                     &     (0.172)         &                     &                     &        (0.174)                 \\

&&&&&&\\
Mean of dep. var.        &        11.3         &        9.33         &          10         &        10.8         &        9.06         &         9.7         \\
Observations        &        2,597         &        4,762         &        7,359         &        3,043         &        5,235         &        8,278         \\
R-squared           &       0.300         &       0.270         &       0.333         &       0.204         &       0.247         &       0.274         \\
 
\bottomrule
\end{tabular}%
\vspace{1mm}

\begin{minipage}{1\linewidth}
 {\footnotesize Note: The table above reports estimates of eq. (\ref{eq:child-parent-gp-regression}) interacted with whether the birthplace of the children (G3) is in a rural area. Standard errors are clustered at the family level. A birthplace is designated to be rural if an individual is born in a village. Birthplace is otherwise designated to be urban if an individual is born in a small town or big city. Due to missing observations on the type of birthplace for 991 individuals in our restricted sample (11\% of the 8,278 sample), we use the type of residence at the time of the first IFLS survey for the individual. 522 individuals (about 6\% of the total sample) with missing birthplace information reported living in a rural area, and 449 (5.7\%) reported living in an urban area.   *** p$<$0.01, ** p$<$0.05, * p$<$0.1.}
\end{minipage}
\end{table}
\doublespacing 

\clearpage
\begin{table}[h]
\centering
\singlespacing
\small
\caption{Robustness of Multigenerational Regression Estimates}\label{tab:multi_lineages}
    \begin{tabular}{lcccccccc}
    \toprule
    &\multicolumn{8}{c}{Dependent Variable: Child's Education}    \\
 \hline
Grandparents     &\multicolumn{4}{c}{Maternal}&\multicolumn{4}{c}{Paternal}    \\
  \cmidrule(lr){2-5} \cmidrule(lr){6-9}
          & (1)   & (2)   & (3)   & (4)   &  (5)   & (6)   & (7)   & (8) \\
    \midrule
Father &       0.378\sym{***}&       0.262\sym{***}&       0.389\sym{***}&       0.265\sym{***}&                     &       0.271\sym{***}&                     &       0.271\sym{***}\\
                    &    (0.0117)         &    (0.0143)         &    (0.0113)         &    (0.0141)         &                     &    (0.0148)         &                     &    (0.0144)         \\
 Mother  &                     &       0.217\sym{***}&                     &       0.223\sym{***}&       0.401\sym{***}&       0.224\sym{***}&       0.415\sym{***}&       0.228\sym{***}\\
                    &                     &    (0.0163)         &                     &    (0.0162)         &    (0.0128)         &    (0.0155)         &    (0.0125)         &    (0.0155)         \\

Grandfather                                  &      0.0714\sym{***}&     0.00656         &                     &                     &      0.0525\sym{***}                 &          -0.0278\sym{*}                          &                     &                     \\
                                                       &    (0.0140)         &    (0.0141)         &                     &                     &        (0.0155)                         &             (0.0153)                           &                     &                     \\

Grandmother                          &                     &                     &      0.0617\sym{***}&     -0.0137         &                     &                     &   0.0180         &     -0.0506\sym{***}\\
                                                     &                     &                     &    (0.0175)         &    (0.0177)         &                     &                     &   (0.0191)         &    (0.0185)         \\
 
Constant            &       7.212\sym{***}&       7.048\sym{***}&       7.242\sym{***}&       7.043\sym{***}&       7.666\sym{***}&       7.045\sym{***}&       7.696\sym{***}&       7.028\sym{***}\\
                    &    (0.0834)         &    (0.0829)         &    (0.0832)         &    (0.0828)         &    (0.0784)         &    (0.0828)         &    (0.0777)         &    (0.0826)         \\
&&&&&&&&\\
Mean of y           &         9.7         &         9.7         &         9.7         &         9.7         &         9.7         &         9.7         &         9.7         &         9.7         \\
Observations        &        8,278         &        8,278         &        8,278         &        8,278         &        8,278         &        8,278         &        8,278         &        8,278         \\
R-squared           &       0.240         &       0.267         &       0.238         &       0.267         &       0.217         &       0.267         &       0.216         &       0.268         \\
    \bottomrule
    \end{tabular}%

    \vspace{2mm}
\begin{minipage}{1.05\linewidth}
 {\footnotesize  \textbf{Note}: The table above reports the estimates of multigenerational correlations in education from eq. (\ref{eq:child-parent-gp-regression}) for different combinations of parents and grandparents education. The dependent variable is the child's years of education (G1). Standard errors are clustered at the family level (all children with the same mother are clustered together). *** p$<$0.01, ** p$<$0.05, * p$<$0.1.}
\end{minipage}

\end{table}

\clearpage
\singlespacing

\begin{table}[H]
\small
\singlespacing
\centering
\caption{Multigenerational Transmission by Province-Level Education Expenditure Share}
\label{tab:area_share}
\begin{tabular}{lcccccc}
\toprule
      & \multicolumn{6}{c}{Dependent Variable: Child's Education} \\
\midrule
      & \multicolumn{3}{c}{Expenditure Share} & \multicolumn{3}{c}{Dummy for most expensive prov.} \\
      & All Children & Male  & Female & All Children & Male  & Female \\
      & (1)   & (2)   & (3) & (4) & (5) & (6) \\
\midrule
Parental Education  &       0.507\sym{***}&       0.496\sym{***}&       0.519\sym{***}&       0.529\sym{***}&       0.515\sym{***}&       0.545\sym{***}\\
                    &    (0.0141)         &    (0.0181)         &    (0.0184)         &    (0.0155)         &    (0.0200)         &    (0.0201)         \\
\ $\times$ Expenditure &      -0.238\sym{**} &      -0.227         &      -0.242\sym{*}  &      -0.181\sym{***}&      -0.156\sym{***}&      -0.207\sym{***}\\
                    &     (0.111)         &     (0.139)         &     (0.141)         &    (0.0364)         &    (0.0454)         &    (0.0469)         \\
&&&\\
Grandparental Education   &     -0.0123         &    -0.00709         &     -0.0193         &      0.0119         &      0.0247         &    -0.00108         \\
                    &    (0.0222)         &    (0.0278)         &    (0.0292)         &    (0.0243)         &    (0.0303)         &    (0.0322)         \\
\ $\times$ Expenditure   &      -0.210         &      -0.326         &      -0.103         &     -0.0885\sym{*}  &      -0.161\sym{**} &     -0.0194         \\
                    &     (0.162)         &     (0.213)         &     (0.205)         &    (0.0483)         &    (0.0656)         &    (0.0597)         \\
&&&&&&\\
Expenditure         &       4.233\sym{***}&       4.075\sym{***}&       4.373\sym{***}&       2.449\sym{***}&       2.403\sym{***}&       2.500\sym{***}\\
                    &     (0.721)         &     (0.937)         &     (0.864)         &     (0.251)         &     (0.325)         &     (0.301)         \\

&&&\\
Observations        &        7,222         &        3,692         &        3,530         &        7,222         &        3,692         &       3,530         \\
R-squared           &       0.334         &       0.320         &       0.350         &       0.342         &       0.330         &       0.357         \\

\bottomrule
\end{tabular}%
\vspace{1mm}
\begin{minipage}{0.95\linewidth}
{\footnotesize \textit{Note:} OLS estimates of eq. (\ref{eq:child-parent-gp-regression}), augmented with an interaction term between the province-level household expenditure share on education and the ancestors' education, alongside the expenditure share's main effect (Columns (1)--(3)) and with an interaction term between a dummy for these expenditure being more than 40\% of household earnings at the province level and the ancestors' education, alongside the dummy's main effect (Columns (4)--(6)). The provinces where we compute educational expenditures being above 40\% of household earnings are DKI Jakarta and DI Yogyakarta. For a definition of the expenditure variable and of the expenditure share see sections \ref{sec:varinfo} and \ref{app:expenditure}. The analysis is restricted to G1 children residing in one of the 13 provinces originally surveyed in the IFLS, ensuring sufficient observations for reliable province-level expenditure share estimates. Standard errors are clustered at the family level. *** $p<0.01$, ** $p<0.05$, * $p<0.1$.}
\end{minipage}
\end{table}

\doublespacing 

\begin{table}[t]
\singlespacing
  \centering
  \caption{Multigenerational Transmission and Assortative Practices for Different Lineages}\label{tab:ass_mat_lin}  %

\begin{tabular}{lccccc}
    \toprule
       &\multicolumn{5}{c}{Dependent Variable: Child's Education} \\
     \hline
    {\it Parent}     &\multicolumn{2}{c}{Father}   &\multicolumn{2}{c}{Mother}   & Average   \\
    {\it Grandparent}     & Paternal   & Maternal   & Maternal    & Paternal   & Average   \\
      
    & (1)   & (2)   & (3)   & (4)   & (5)   \\

    \midrule
    Parental Education       &       0.420\sym{***}&       0.392\sym{***}&       0.408\sym{***}&       0.411\sym{***}&       0.514\sym{***}\\
                    &    (0.0136)         &    (0.0136)         &    (0.0160)         &    (0.0146)         &    (0.0162)         \\
    Parental Education     & -0.0616\sym{**} &     -0.0552\sym{*}  &     -0.0260         &     -0.0198         &     -0.0300         \\
\ $\times 1(FamilyChoice)$  &    (0.0287)         &    (0.0289)         &    (0.0344)         &    (0.0311)         &    (0.0381)         \\
      
              &       &       &       &       &         \\
  Grandparental Education             &     -0.0150         &      0.0713\sym{***}&      0.0319         &      0.0307         &     -0.0673\sym{**} \\
                    &    (0.0218)         &    (0.0206)         &    (0.0223)         &    (0.0209)         &    (0.0264)         \\
    Grandparental Education            &     0.0987\sym{*}  &      0.0806\sym{*}  &      0.0753\sym{*}  &      0.0873\sym{*}  &       0.117\sym{*}  \\
     \ $\times 1(FamilyChoice)$ &    (0.0532)         &    (0.0428)         &    (0.0429)         &    (0.0523)         &    (0.0604)         \\
     
           &       &       &       &       &         \\
    $1(FamilyChoice)$           &       0.338\sym{*}  &       0.314\sym{*}  &     -0.0136         &     -0.0142         &       0.150         \\
                    &     (0.174)         &     (0.175)         &     (0.165)         &     (0.165)         &     (0.176)         \\ 
           &       &       &       &       &         \\
    Constant                &       7.121\sym{***}&       7.114\sym{***}&       7.682\sym{***}&       7.675\sym{***}&       6.983\sym{***}\\
                    &     (0.105)         &     (0.105)         &    (0.0970)         &    (0.0979)         &     (0.103)         \\
&&&&&\\
 Observations        &        8,161         &        8,161         &        8,161         &        8,161         &        8,161         \\
R-squared           &       0.237         &       0.241         &       0.218         &       0.218         &       0.268         \\
    \bottomrule
    \end{tabular}%
        \vspace{2mm}

\begin{minipage}{0.86\linewidth}
 \footnotesize{Note: The table reports the estimated parameters of different variants of eq. (\ref{eq:child-parent-gp-regression}). The indicator \textit{FamilyChoice} equals 1 if the mother declares that her husband was selected by her family. Grandparental education is the average of the years of education completed by the parents of the father (columns 1 and 4) or the mother (columns 2 and 3). Each regression controls for birth cohort of the wife fixed effects, standard errors are clustered at the family level. *** p$<$0.01, ** p$<$0.05, * p$<$0.1}. 
\end{minipage}

\end{table}

\begin{table}[htbp]
\singlespacing
\centering
\caption{Crisis Impact on Multigenerational Links in Education (Table 4 cont.) \label{tab:did_crisis_app}}
\begin{tabular}{lccc@{\hspace{1em}}ccc}
\hline
 & (1) & (2) & (3) & (4) & (5) & (6) \\
 & \multicolumn{3}{c}{Specification 1} & \multicolumn{3}{c}{Specification 2} \\
 & \multicolumn{3}{c}{Pooling treated cohorts} & \multicolumn{3}{c}{Separating treated cohorts} \\
  \cmidrule(lr){2-4}\cmidrule(lr){5-7}  
Sample & All & Males & Females & All & Males & Females \\
\hline
Born 79--88 & 0.948*** & 0.707*** & 1.206*** &  &  &  \\
 & (0.1470) & (0.1820) & (0.2130) &  &  &  \\
Born 79--82 (Upper secondary age) &  &  &  & 0.689*** & 0.538*** & 0.946*** \\
 &  &  &  & (0.1630) & (0.1730) & (0.2330) \\
Born 83--85 (Lower secondary age) &  &  &  & 1.115*** & 0.569** & 1.672*** \\
 &  &  &  & (0.1940) & (0.2530) & (0.2690) \\
Born 86--88 (Primary school age) &  &  &  & 1.247*** & 1.105*** & 1.476*** \\
 &  &  &  & (0.2500) & (0.2650) & (0.2990) \\
Born 79--88 x crisis & 0.0729 & 0.0785 & 0.0198 &  &  &  \\
 & (0.1020) & (0.1360) & (0.1780) &  &  &  \\
Born 79--82 x crisis &  &  &  & -0.1560 & -0.0172 & -0.2670 \\
 &  &  &  & (0.1660) & (0.1590) & (0.2530) \\
Born 83--85 x crisis &  &  &  & 0.2350 & 0.0282 & 0.2870 \\
 &  &  &  & (0.1620) & (0.2570) & (0.3410) \\
Born 86--88 x crisis &  &  &  & 0.0165 & 0.1180 & -0.1120 \\
 &  &  &  & (0.1590) & (0.2600) & (0.2440) \\
Parent edu & 0.501*** & 0.466*** & 0.535*** & 0.521*** & 0.490*** & 0.552*** \\
 & (0.0383) & (0.0379) & (0.0473) & (0.0425) & (0.0390) & (0.0574) \\
Parent edu x crisis & -0.0174 & -0.0288 & -0.0022 & -0.0248 & -0.0478 & 0.0016 \\
 & (0.0372) & (0.0440) & (0.0452) & (0.0414) & (0.0465) & (0.0556) \\
Parent edu x Born 79--88 & -0.138*** & -0.0877** & -0.186*** &  &  &  \\
 & (0.0332) & (0.0409) & (0.0399) &  &  &  \\
Parent edu x Born 79--82 &  &  &  & -0.103*** & -0.0840** & -0.130** \\
 &  &  &  & (0.0291) & (0.0338) & (0.0493) \\
Parent edu x Born 83--85 &  &  &  & -0.158*** & -0.0638 & -0.242*** \\
 &  &  &  & (0.0414) & (0.0576) & (0.0583) \\
Parent edu x Born 86--88 &  &  &  & -0.204*** & -0.190*** & -0.221*** \\
 &  &  &  & (0.0555) & (0.0658) & (0.0570) \\
Parent edu x Born 79--88 x crisis & -0.0108 & -0.0010 & -0.0239 &  &  &  \\
 & (0.0354) & (0.0398) & (0.0434) &  &  &  \\
Parent edu x Born 79--82 x crisis &  &  &  & 0.0211 & 0.0237 & 0.0165 \\
 &  &  &  & (0.0360) & (0.0356) & (0.0559) \\
Parent edu x Born 83--85 x crisis &  &  &  & -0.0184 & 0.0461 & -0.0679 \\
 &  &  &  & (0.0409) & (0.0521) & (0.0677) \\
Parent edu x Born 86--88 x crisis &  &  &  & -0.0099 & 0.0031 & -0.0308 \\
 &  &  &  & (0.0449) & (0.0514) & (0.0498) \\
Constant & 6.924*** & 7.002*** & 6.847*** & 6.815*** & 6.939*** & 6.653*** \\
 & (0.1990) & (0.1870) & (0.2530) & (0.2250) & (0.2060) & (0.2870) \\
\hline
Observations & 7,247 & 3,710 & 3,537 & 7,247 & 3,710 & 3,537 \\
R-squared & 0.3570 & 0.3500 & 0.3950 & 0.3600 & 0.3530 & 0.4000 \\
\hline
\end{tabular}

\vspace{0.5em}
\parbox{\textwidth}{\footnotesize
\textit{Notes:} See \autoref{tab:did_crisis_main}.}
\end{table}

\singlespacing

\begin{table}[htbp]
\centering
\caption{Crisis Impact on Multigenerational Links in Education: Placebo Test \label{tab:did_crisis_placebo}}
\begin{tabular}{lccc}
\hline
 & (1) & (2) & (3) \\
Sample & All & Males & Females \\
\hline
GP edu & 0.0328 & 0.0541 & 0.0136 \\
 & (0.0729) & (0.0874) & (0.0988) \\
GP edu x crisis & 0.0226 & 0.0799 & -0.0558 \\
 & (0.0414) & (0.0586) & (0.0847) \\
Born 77--78 x GP edu & 0.0078 & -0.0562 & 0.0665 \\
 & (0.0667) & (0.1050) & (0.1240) \\
Born 77--78 x GP edu x crisis & -0.0532 & 0.0009 & -0.0969 \\
 & (0.0908) & (0.1190) & (0.2030) \\
Parent edu & 0.462*** & 0.426*** & 0.534*** \\
 & (0.0557) & (0.0633) & (0.0637) \\
Parent edu x crisis & -0.0299 & -0.0939* & 0.0032 \\
 & (0.0397) & (0.0524) & (0.0555) \\
Born 77--78 x Parent edu & 0.0291 & 0.0494 & -0.0364 \\
 & (0.0494) & (0.0876) & (0.0452) \\
Born 77--78 x Parent edu x crisis & 0.0049 & 0.0475 & -0.0143 \\
 & (0.0560) & (0.0701) & (0.0876) \\
Born 77--78 & -0.2210 & -0.2300 & 0.0210 \\
 & (0.3310) & (0.5030) & (0.3430) \\
Born 77--78 x crisis & -0.0569 & -0.3110 & -0.0162 \\
 & (0.3370) & (0.4070) & (0.3870) \\
Constant & 7.190*** & 7.281*** & 6.958*** \\
 & (0.2240) & (0.2820) & (0.2830) \\
\hline
Observations & 1,676 & 921 & 755 \\
R-squared & 0.4850 & 0.4810 & 0.5750 \\
\hline
\end{tabular}

\vspace{0.5em}
\parbox{\textwidth}{\footnotesize
\textit{Notes:} Sample of analysis is restricted to those individuals born between 1975 and 1978. OLS regression of years of education on average parental education and average grandparental education, fully interacted with an indicator for whether the child was born in 1977 or 1978 (as opposed to 1975 or 1976) and the variable ``crisis'', which measures exposure to the crisis as the proportionate change in nighttime lights in the kabupaten of residence in 1993 between 1996 and 1998. This measure is then standardised to have mean 0 and standard deviation 1 in the analysis sample. All specifications also control for kabupaten fixed effects and cluster the standard errors at the Karesidenan level, which is a geographical unit larger than kabupaten but smaller than province. *** $p<0.01$, ** $p<0.05$, * $p<0.1$.}
\end{table}
\singlespacing

\begin{table}[htbp]
\centering
\caption{Crisis Impact on Multigenerational Links in Education: Robustness Checks (Male sample)
\label{tab:did_crisis_robust}}
\begin{tabular}{lccccc}
\hline
 & (1) & (2) & (3) & (4) & (5) \\
Outcome & \multicolumn{3}{c}{Years of education} & Edu $\geq$ 9 yrs & Edu $\geq$ 12 years \\
\cmidrule(lr){2-4} \cmidrule(lr){5-6}  
Clustering of SE & Karesidenan & Kabupaten & Province & Karesidenan & Karesidenan \\
\hline
GP edu & -0.0221 & -0.0221 & -0.0221 & -0.0097 & 0.0108 \\
 & (0.0625) & (0.0563) & (0.0542) & (0.0074) & (0.0078) \\
GP edu x crisis & 0.0689 & 0.0689 & 0.0689 & 0.0130 & 0.0096 \\
 & (0.0579) & (0.0562) & (0.0571) & (0.0078) & (0.0069) \\
GP edu x Born 79--82& 0.0556 & 0.0556 & 0.0556 & 0.0145** & -0.0095 \\
 & (0.0612) & (0.0692) & (0.0475) & (0.0061) & (0.0120) \\
GP edu x Born 83--85  & 0.0266 & 0.0266 & 0.0266 & -0.0020 & -0.0099 \\
 & (0.0703) & (0.0811) & (0.0671) & (0.0108) & (0.0132) \\
GP edu x Born 86--88  & 0.0227 & 0.0227 & 0.0227 & 0.0083 & 0.0028 \\
 & (0.0886) & (0.0830) & (0.1000) & (0.0118) & (0.0100) \\
GP edu x Born 79--82 x crisis & -0.1190** & -0.1190 & -0.1190** & -0.0182** & -0.0187** \\
 & (0.0452) & (0.0719) & (0.0438) & (0.0083) & (0.0087) \\
GP edu x Born 83--85 x crisis & -0.1490** & -0.1490** & -0.1490** & -0.0250** & -0.0042 \\
 & (0.0643) & (0.0746) & (0.0652) & (0.0095) & (0.0107) \\
GP edu x Born 86--88 x crisis & -0.0753 & -0.0753 & -0.0753 & -0.0100 & -0.0102 \\
 & (0.0729) & (0.0734) & (0.0743) & (0.0110) & (0.0099) \\
 & & & & & \\
Parent edu & 0.4900*** & 0.4900*** & 0.4900*** & 0.0615*** & 0.0380*** \\
 & (0.0390) & (0.0359) & (0.0392) & (0.0051) & (0.0046) \\
Parent edu x crisis & -0.0478 & -0.0478 & -0.0478 & -0.0050 & 0.0024 \\
 & (0.0465) & (0.0391) & (0.0510) & (0.0064) & (0.0046) \\
Parent edu x Born 79--82& -0.0840** & -0.0840** & -0.0840** & -0.0103** & 0.0020 \\
 & (0.0338) & (0.0387) & (0.0338) & (0.0042) & (0.0054) \\
Parent edu x Born 83--85  & -0.0638 & -0.0638 & -0.0638 & -0.0076 & -0.0002 \\
 & (0.0576) & (0.0542) & (0.0519) & (0.0062) & (0.0102) \\
Parent edu x Born 86--88  & -0.1900*** & -0.1900*** & -0.1900** & -0.0242*** & -0.0074 \\
 & (0.0658) & (0.0585) & (0.0624) & (0.0066) & (0.0064) \\
Parent edu x Born 79--82 x crisis & 0.0237 & 0.0237 & 0.0237 & 0.0048 & 0.0074* \\
 & (0.0356) & (0.0391) & (0.0409) & (0.0057) & (0.0039) \\
Parent edu x Born 83--85 x crisis & 0.0461 & 0.0461 & 0.0461 & 0.0104 & 0.0040 \\
 & (0.0521) & (0.0546) & (0.0555) & (0.0076) & (0.0085) \\
Parent edu x Born 86--88 x crisis & 0.0031 & 0.0031 & 0.0031 & 0.0017 & -0.0011 \\
 & (0.0514) & (0.0487) & (0.0645) & (0.0072) & (0.0061) \\

\hline
Observations & 3,710 & 3,710 & 3,710 & 3,710 & 3,710 \\
R-squared & 0.3530 & 0.3530 & 0.3530 & 0.3170 & 0.2470 \\
\hline
\end{tabular}

\vspace{0.5em}
\parbox{\textwidth}{\footnotesize
\textit{Notes:} OLS regression on the male sample of years of education on average parental education and average grandparental education, fully interacted with binned cohort dummies and the variable ``crisis'', which measures exposure to the crisis as the proportionate change in nighttime lights in the kabupaten of residence in 1993 between 1996 and 1998. This measure is then standardised to have mean 0 and standard deviation 1 in the analysis sample. All specifications also control for kabupaten fixed effects and cluster the standard errors at the Karesidenan level, which is a geographical unit larger than kabupaten but smaller than province. Cohorts are binned such that the potentially affected cohorts are split into 3 depending on whether they were in upper secondary school (born 79--82), lower secondary school (born 83--85), or primary school (born 86--88) in 1997 when the crisis hit Indonesia. In Column 1, we cluster standard errors at the karesidenan level (same as Table 4). In Column 2, we cluster SEs at the kabupaten level and in Column 3 at the province level. In Columns 4 and 5, we replace the dependent variable by a dummy which takes the value 1 if years of education are at least 9 years (end of lower secondary) and 12 years (end of upper secondary), respectively. *** $p<0.01$, ** $p<0.05$, * $p<0.1$.}
\end{table}

\clearpage

\begin{figure}[h!]\centering
        \caption{Years of Education, Separately for Three Generations}\label{fig:distr_yoe_all}
                \centering
                \includegraphics[width=1.1\textwidth]{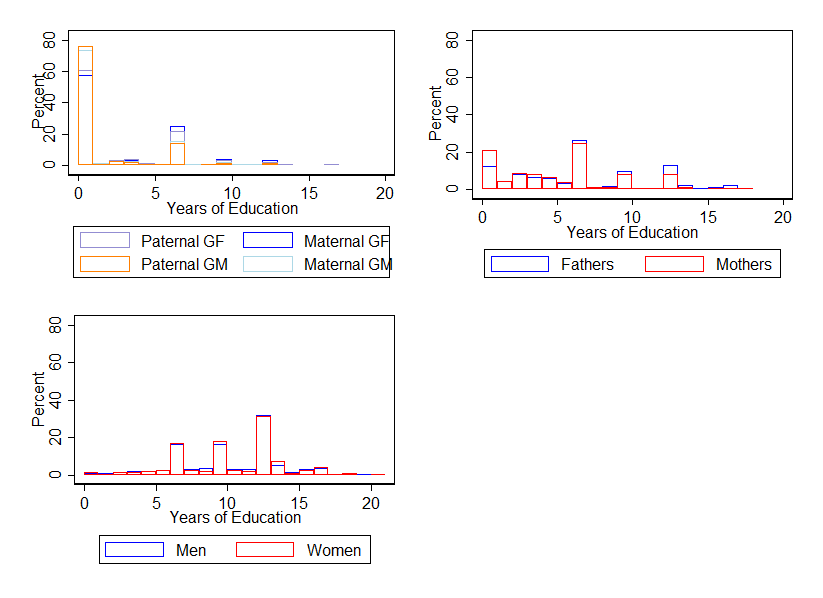}
\begin{minipage}{1.0\textwidth} \setstretch{1}
{\footnotesize \textbf{Note}: The panels show the distribution of education among the grandparents, parents, and the grandchildren in the IFLS sample.}
\end{minipage}
\end{figure}

\clearpage
 
\begin{figure}[h]\centering
        \caption{Education by Cohort for Parents in IFLS and Census}\label{fig:validation}
                \centering
                \includegraphics[width=0.8\textwidth]{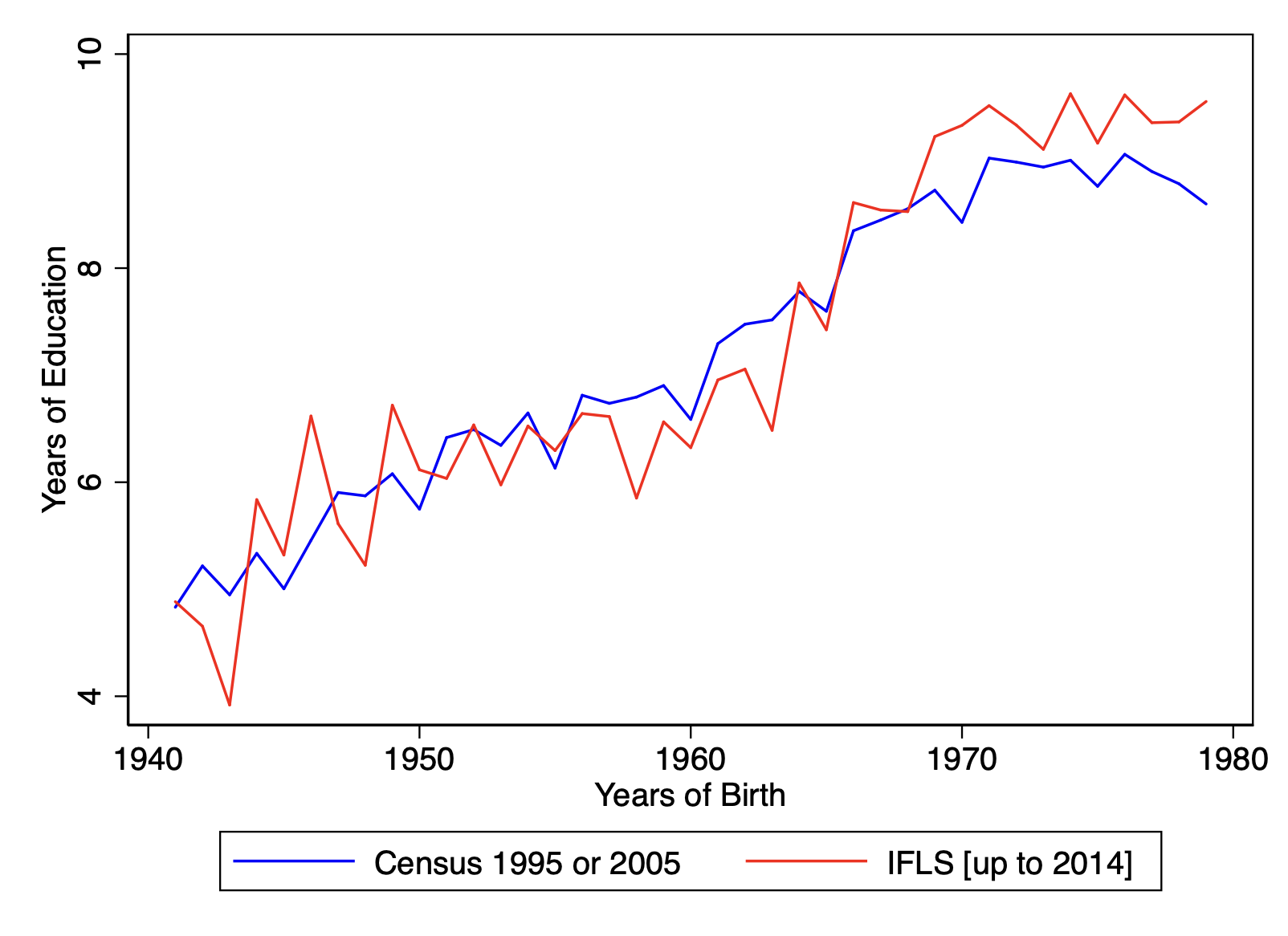}
\begin{minipage}{1.0\textwidth} \setstretch{1}
{\footnotesize \textbf{Note}: The figure plots the average years of education per cohort (born between 1940 to 1980) in the 1995 or 2005 Census and in the IFLS panel.}
\end{minipage}
\end{figure}

\begin{figure}[h]\centering
        \caption{Education by Cohort for Grandparents in IFLS and Census}\label{fig:validation_gp}
                \centering
                \includegraphics[width=0.8\textwidth]{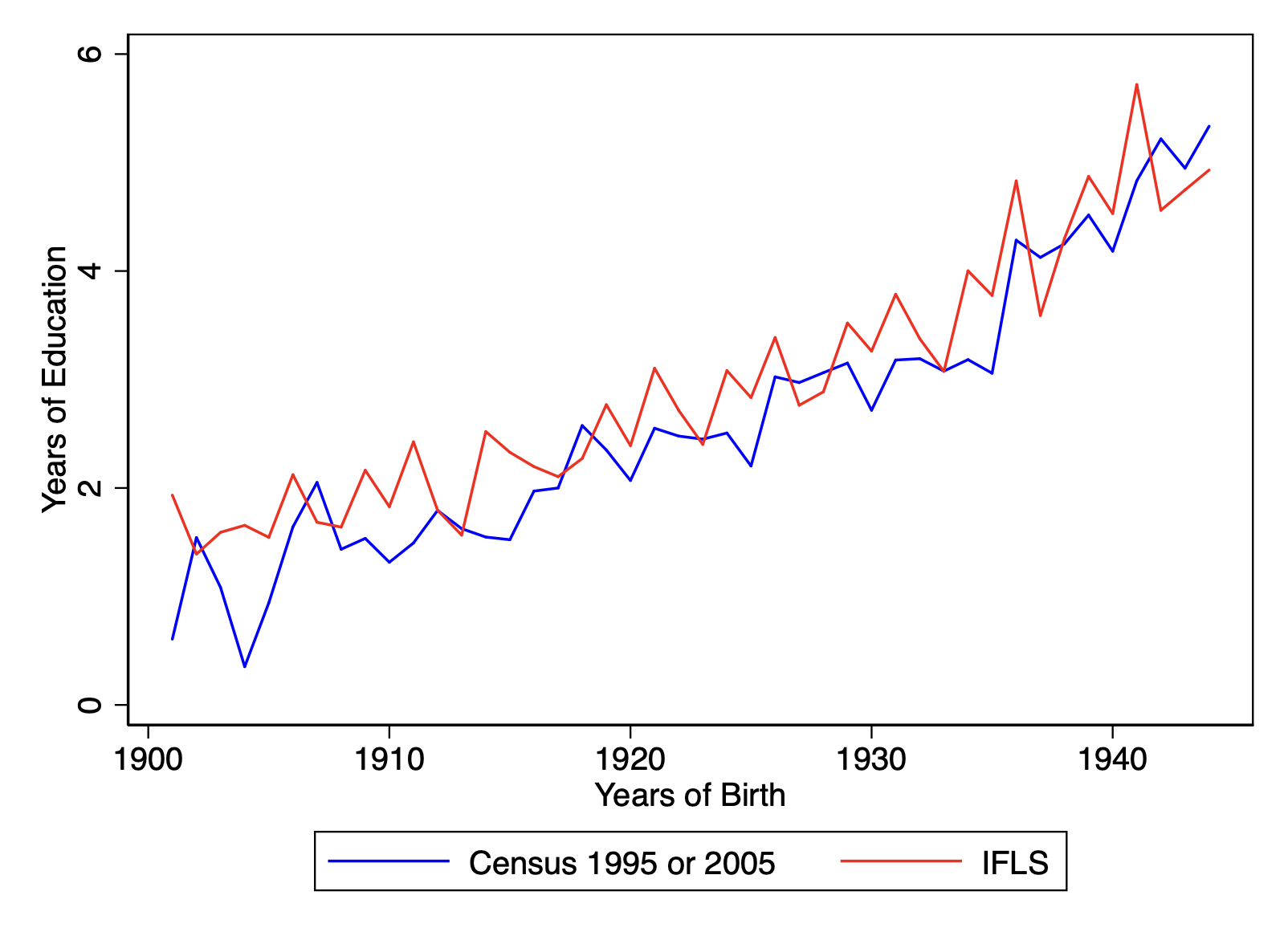}\\
                \includegraphics[width=0.8\textwidth]{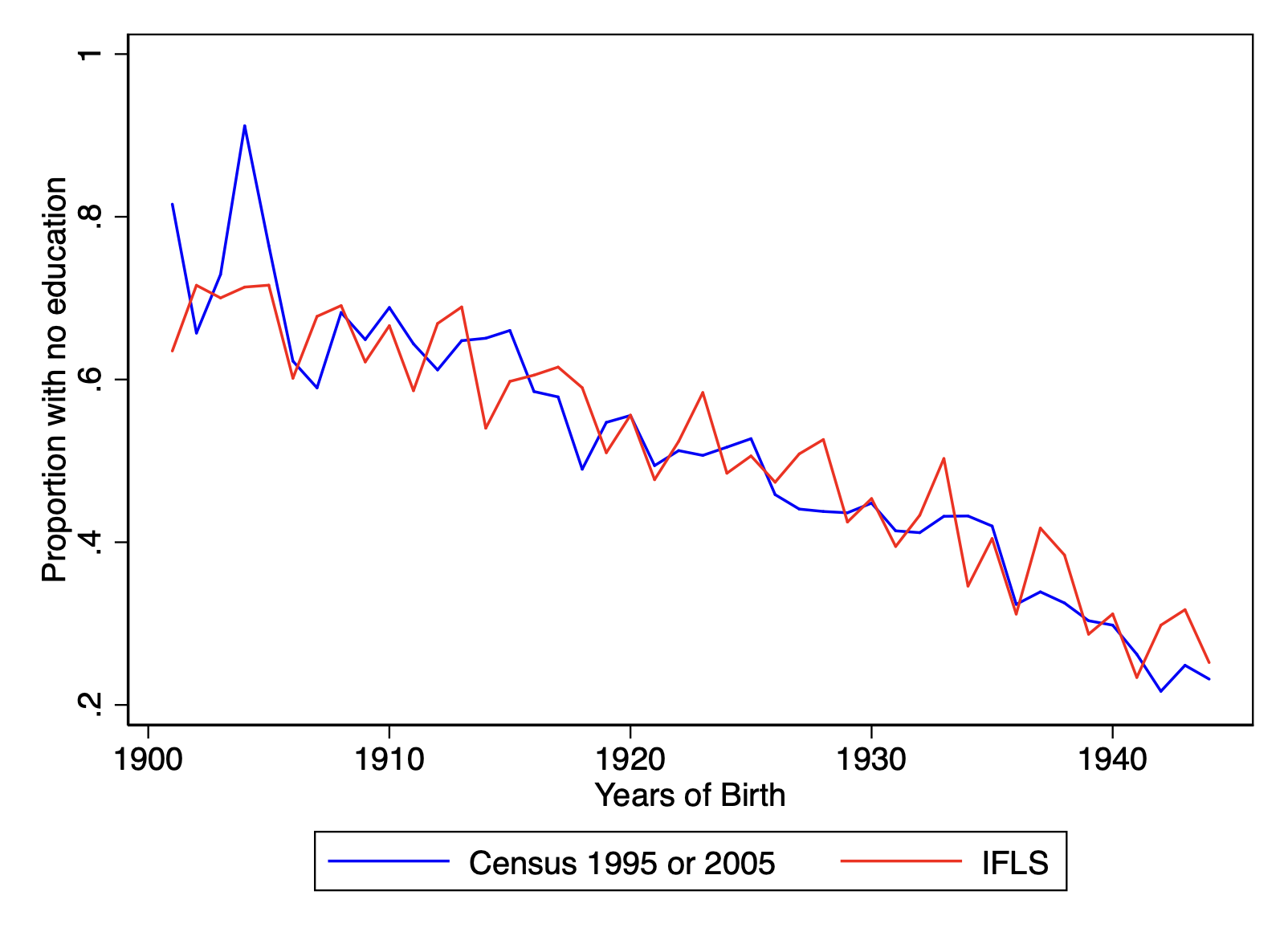}
\begin{minipage}{1.0\textwidth} \setstretch{1}
{\footnotesize \textbf{Note}: The top panel plots the average years of education per cohort (born between 1900 to 1940) in the 1995 or 2005 Census and in the IFLS panel. The bottom panel plots the proportion of respondents with no education per cohort (born between 1900 to 1940) in the 1995 or 2005 Census and in the IFLS panel.}
\end{minipage}
\end{figure}

\begin{figure}[h]\centering
        \caption{Years of Birth, Cohorts 1975-1988}\label{fig:distr_yob}
                \centering
                \includegraphics[width=0.75\textwidth]{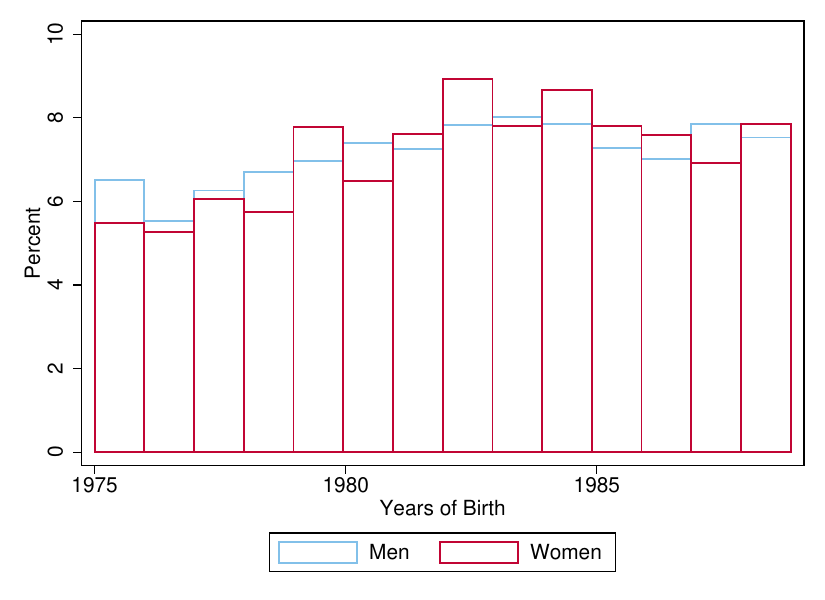}
\begin{minipage}{1.0\textwidth} \setstretch{1}
{\footnotesize \textbf{Notes}: The figure plots the distribution of birth cohorts in our grandchildren sample by gender.}
\end{minipage}
\end{figure}

\begin{figure}[h!]\centering
        \caption{Distribution of Years of Birth, Parents}\label{fig:distr_yob_p}
                \centering
                \includegraphics[width=0.75\textwidth]{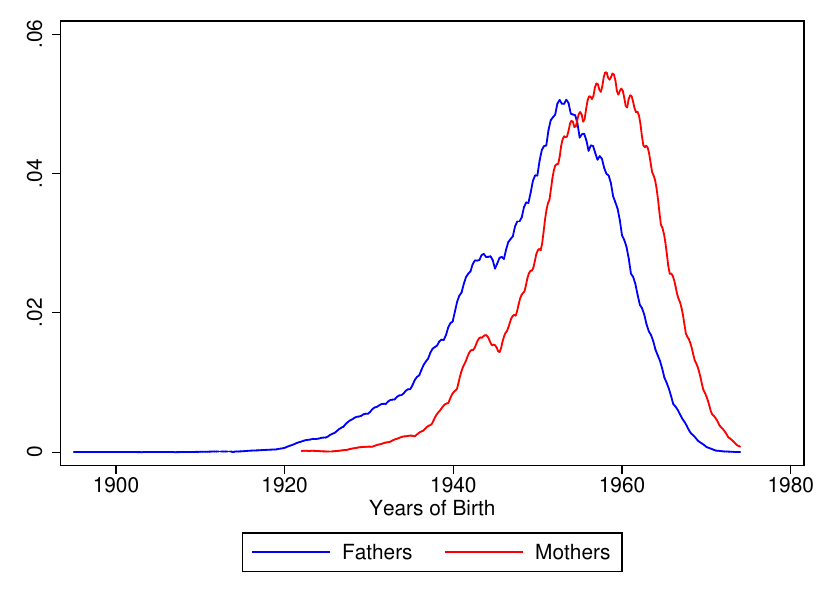}
\begin{minipage}{1.0\textwidth} \setstretch{1}
{\footnotesize \textbf{Note}: Kernel density estimation for the distribution of years of birth of fathers and mothers of individuals born between 1975 and 1988.}
\end{minipage}
\end{figure}

\begin{figure}[h!]\centering
        \caption{Distribution of Years of Birth, Grandparents}\label{fig:distr_yob_gp}
                \centering
                \includegraphics[width=0.8\textwidth]{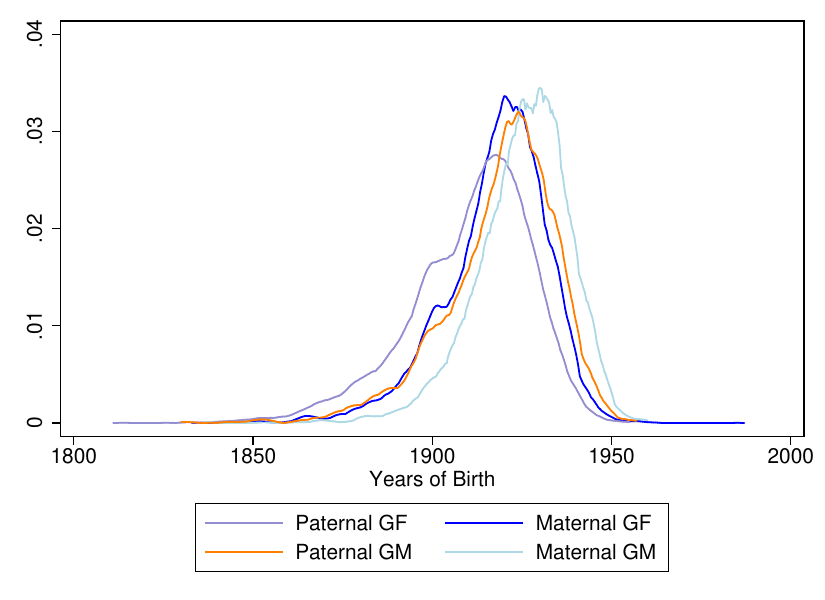}
\begin{minipage}{1.0\textwidth} \setstretch{1}
{\footnotesize \textbf{Note}: Kernel density estimation for the distribution of years of birth of grandfathers and grandmothers by lineage.}
\end{minipage}
\end{figure}

\begin{figure}[h]\centering
        \caption{Distribution of Years of Education by Generation in the IFLS}\label{fig:distr_yoe3}
                \centering
                \includegraphics[width=0.8\textwidth]{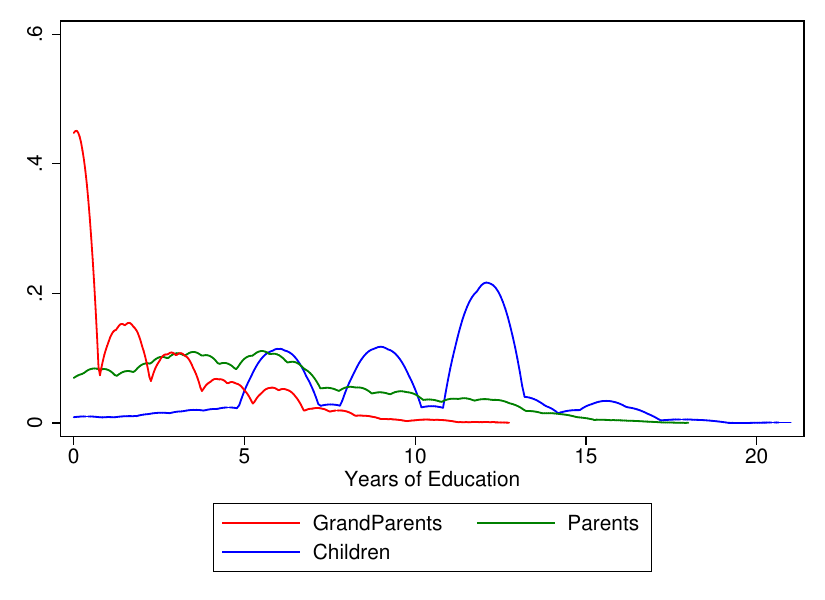}
\begin{minipage}{1.0\textwidth} \setstretch{1}
{\footnotesize \textbf{Note}: Kernel density estimation for the distribution of years of education of all three generations.}
\end{minipage}
\end{figure}

\begin{figure}[h]\centering
        \caption{Children (G1) Average Years of Education as a Function of Average Grandparents' Years of Education, separately for Parents with Different Education Levels}\label{fig:multigen_plot}
                \centering
                \includegraphics[width=0.8\textwidth]{Figures/edu_m_gp_contrp.png}
                
\begin{minipage}{0.8\textwidth} \setstretch{1}
{\footnotesize \textbf{Note}: The figure plots the average years of education of children (y-axis) given the years of grandparents (computed as the average of all four grandparents, x-axis), separately for different education level of parents (No edu=both parents have zero education; Parents primary=avg. parental years of education between 1 and 6; Parents secondary=avg. parental years of education between 7 and 9; Parents $>$ secondary=avg. parental years of education above 9).}
\end{minipage}
\end{figure}

\begin{figure}[h]\centering
        \caption{Estimated Returns to Education by Province}\label{fig:rte}
                \centering
                \includegraphics[width=0.8\textwidth]{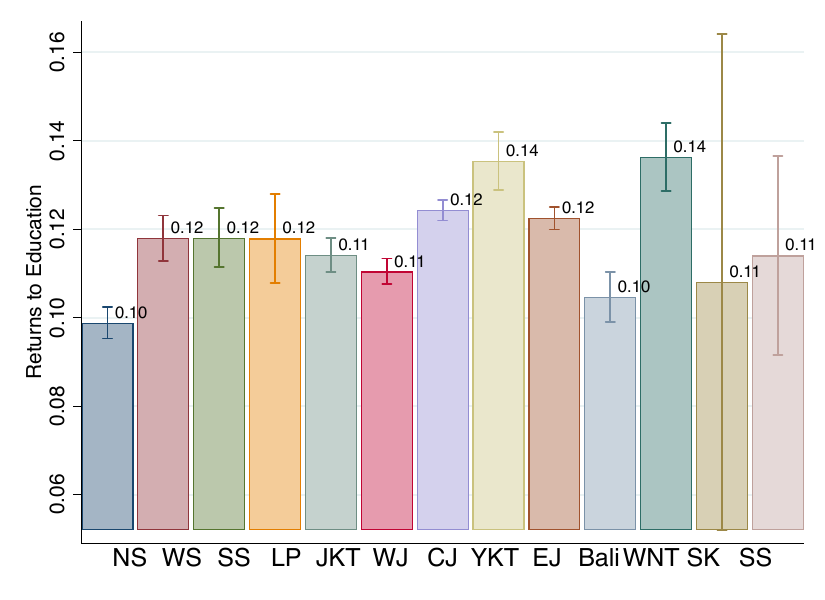}
                
\begin{minipage}{0.8\textwidth} \setstretch{1}
{\footnotesize \textbf{Note}: The figure plots the estimated returns to education at the province level using information from the 1995 Census. The sample is restricted to individuals born before 1975. NS refers to North Sumatra, WS refers to West Sumatra, SS refers to South Sumatra, LP refers to Lampung, JKT is DKI Jakarta, WJ is West Java, CJ is Central Java, YKT is DI Yogyakarta, EJ is East Java, WNT is West Nusa Tenggara, SK refers to South Kalimantan, and SS refers to South Sulawesi.}
\end{minipage}
\end{figure}

\newgeometry{left=0.6cm,right=0.6cm}
\begin{figure}
\caption{Validating Nighttime Lights as a Measure of Crisis Exposure \label{fig:did_crisis_validation}}
    \centering
   \subfigure[Variation in IFLS sample]{\includegraphics[width=0.45\linewidth]{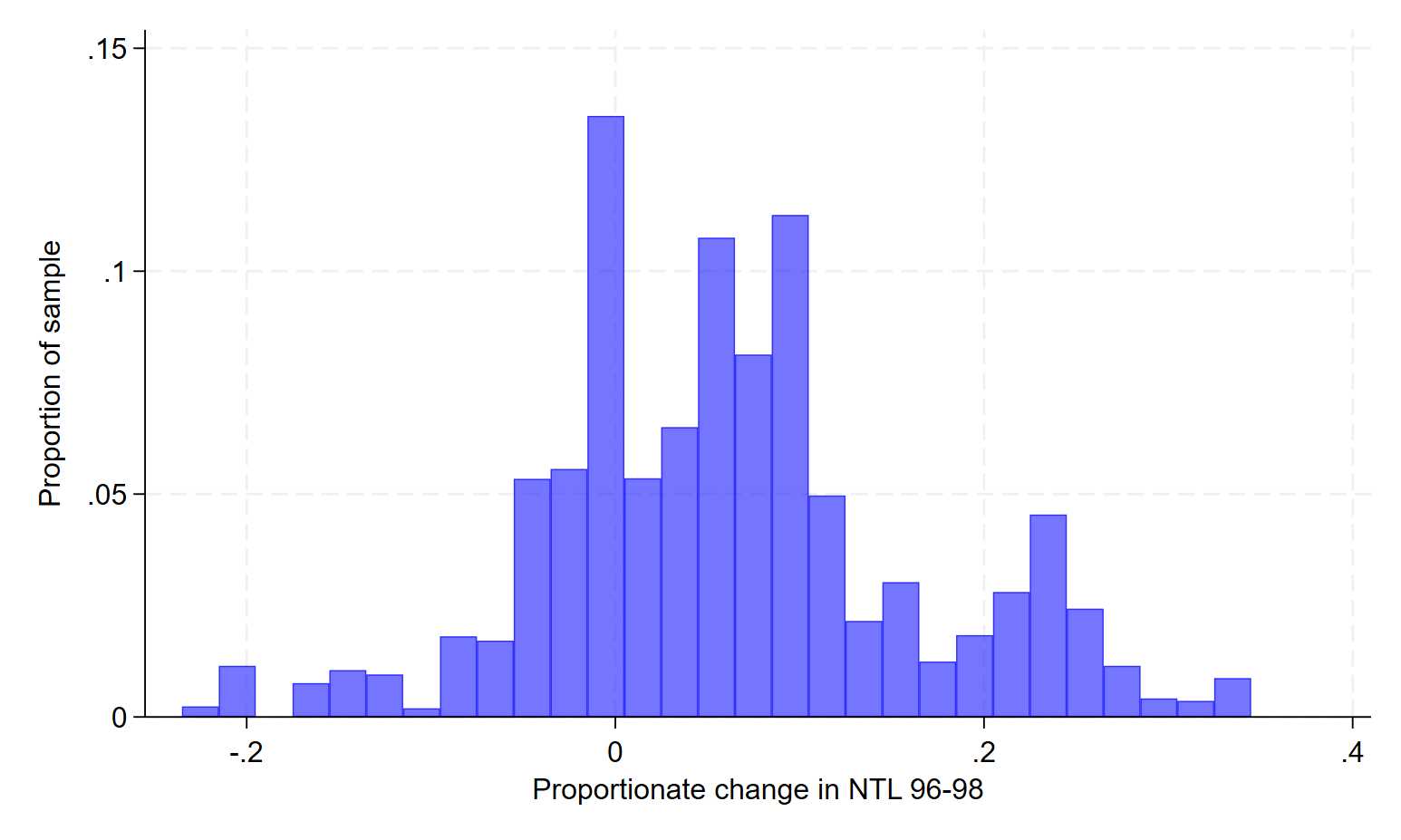}}
   \subfigure[Variation in IFLS sample across urban and rural kabupaten]{\includegraphics[width=0.45\linewidth]{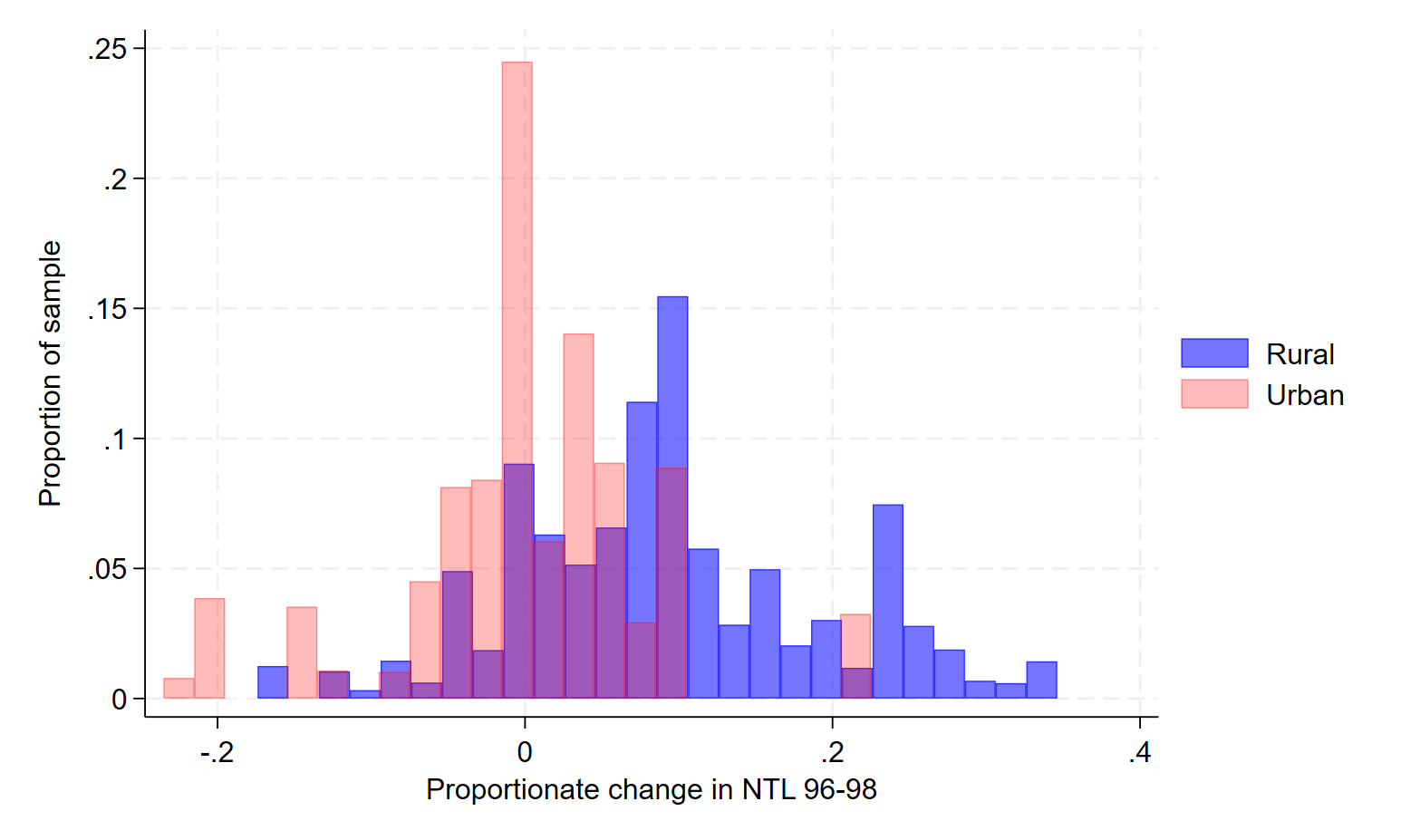}} \\
   \subfigure[Co-variation with province level change in GDP 96-98]{\includegraphics[width=0.45\linewidth]{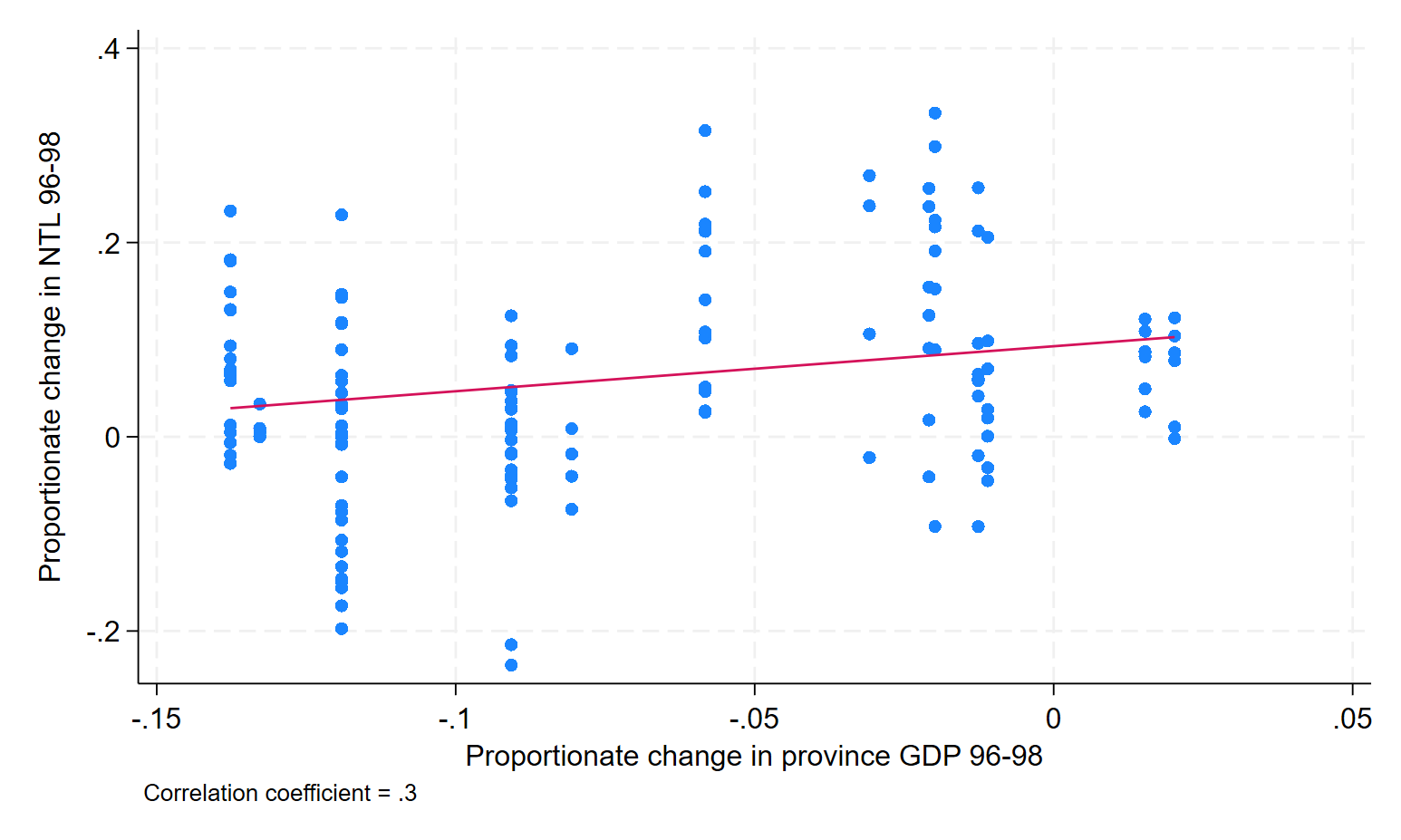}}
   \subfigure[Co-variation with kabupaten-level predicted change in GDP 96-98]{\includegraphics[width=0.45\linewidth]{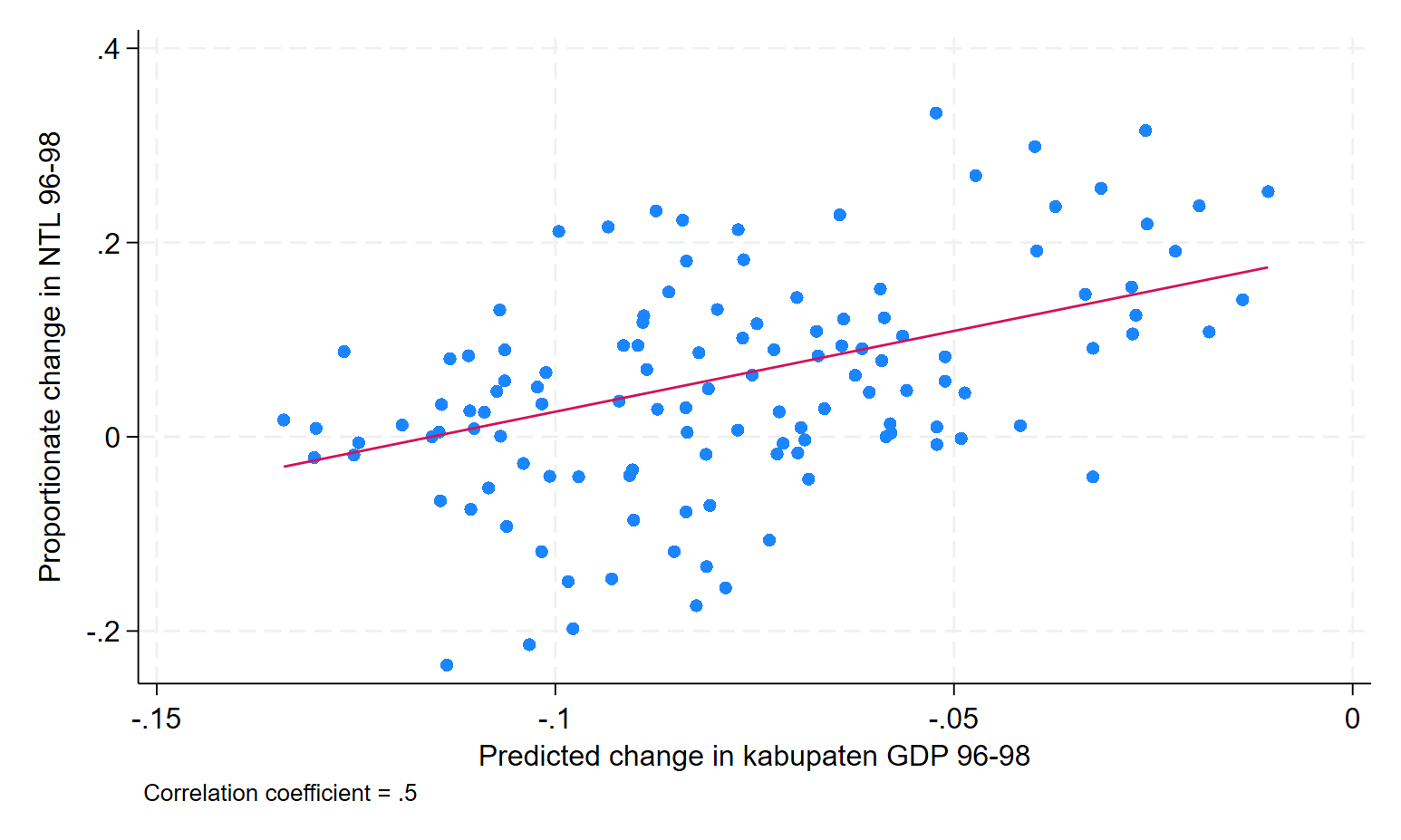}}
\begin{minipage}{1\textwidth} \setstretch{0.8}
{\footnotesize \textbf{Note}: Subfigure (a) is a histogram of the proportionate change in nighttime light intensity (measured on a scale 0-63) between 1996 and 1998 relative to the 1996 value in the kabupaten of residence in 1993 of the IFLS sample. Subfigure (b) is a histogram of the same variable, splitting the sample between kabupatens that are predominantly rural and predominantly urban. Subfigure (c) is a scatter plot of this variable against the proportionate change in province-level GDP between 1996 and 1998. Subfigure (d) is a scatter plot of this variable against the predicted change in kabupaten-level GDP between 1996 and 1998. We construct this Bartik-style predictor of economic shock as a weighted average of the change in GDP in 9 one-digit industries between 1996 and 1998 weighted by the employment share in each of the 8 industries in 1995. The industry-specific GDP figures are taken from the 1996 and 1998 Indonesia's Statistical Yearbooks (Statistik Indonesia), which reports gross domestic product by industrial origin at constant 1993 prices \citep{bps_statistik_indonesia_1996, bps_statistik_indonesia_1998}. The employment shares are constructed using microdata from the 1995 Intercensal Population Survey (SUPAS), obtained form IPUMs International, which provides harmonised census and survey data for Indonesia \citep{ipums_supas_1995}.}
\end{minipage}
\end{figure}\restoregeometry
\end{document}